\documentclass[12pt]{article}
\usepackage{a4wide}
\usepackage{latexsym}
\usepackage{amsmath}
\usepackage{amsfonts}
\usepackage{amscd}
\usepackage{cite}

\usepackage{pslatex}
\usepackage{graphicx}
\usepackage[latin1]{inputenc}
\usepackage[T1]{fontenc}

\allowdisplaybreaks

\newcommand{\bq}{\begin{eqnarray}}
\newcommand{\eq}{\end{eqnarray}}
\newcommand{\eps}{\varepsilon}

\begin{document}

\thispagestyle{empty}

\begin{flushright}
  MITP/15-024 \\
  MaPhy-AvH/2015-03
\end{flushright}

\vspace{1.5cm}

\begin{center}
  {\Large\bf The two-loop sunrise integral around four space-time dimensions
    and generalisations of the Clausen and Glaisher functions towards the elliptic case\\
  }
  \vspace{1cm}
  {\large Luise Adams ${}^{a}$, Christian Bogner ${}^{b}$ and Stefan Weinzierl ${}^{a}$\\
  \vspace{1cm}
      {\small ${}^{a}$ \em PRISMA Cluster of Excellence, Institut f{\"u}r Physik, }\\
      {\small \em Johannes Gutenberg-Universit{\"a}t Mainz,}\\
      {\small \em D - 55099 Mainz, Germany}\\
  \vspace{2mm}
      {\small ${}^{b}$ \em Institut f{\"u}r Physik, Humboldt-Universit{\"a}t zu Berlin,}\\
      {\small \em D - 10099 Berlin, Germany}\\
  } 
\end{center}

\vspace{2cm}

\begin{abstract}\noindent
  {
We present the result for the finite part of the two-loop sunrise integral 
with unequal masses in four space-time dimensions in terms 
of the ${\mathcal O}(\eps^0)$-part and the ${\mathcal O}(\eps^1)$-part
of the sunrise integral around two space-time dimensions.
The latter two integrals are given in terms of elliptic generalisations
of Clausen and Glaisher functions.
Interesting aspects of the result for the ${\mathcal O}(\eps^1)$-part
of the sunrise integral around two space-time dimensions 
are the occurrence of depth two elliptic objects and the weights of the individual terms.
   }
\end{abstract}

\vspace*{\fill}

\newpage

\section{Introduction}
\label{sec:intro}

The two-loop sunrise integral with non-vanishing internal masses is the first and simplest integral in quantum field theory, which
cannot be expressed in terms of multiple polylogarithms.
It has already received considerable attention in the 
literature \cite{Broadhurst:1993mw,Berends:1993ee,Bauberger:1994nk,Bauberger:1994by,Bauberger:1994hx,Caffo:1998du,Laporta:2004rb,Groote:2005ay,Groote:2012pa,Bailey:2008ib,MullerStach:2011ru,Adams:2013nia,Bloch:2013tra,Remiddi:2013joa,Adams:2014vja,Caffo:2002ch,Pozzorini:2005ff,Caffo:2008aw}.
However, the question which generalisations of multiple polylogarithms appear in the evaluation of the two-loop sunrise
integral is still open.
A first and partial answer has been given for the two-loop sunrise integral in two space-time dimensions \cite{Bloch:2013tra,Adams:2014vja}.
In two space-time dimensions the two-loop sunrise integral is finite and does not require regularisation. 
Furthermore only one graph polynomial contributes in the Feynman parameter representation.
It turns out that in two space-time dimensions the two-loop sunrise integral is given
as a product of a period with a sum of elliptic dilogarithms.
This has been shown for the equal mass case in \cite{Bloch:2013tra} and for the general unequal mass case in \cite{Adams:2014vja}.
The arguments of the elliptic dilogarithms have a nice geometric interpretation as the images 
in the Jacobi uniformisation of the intersection points of the variety defined by
the graph polynomial with the integration region.

In this paper we study the two-loop sunrise integral around four space-time dimensions.
Our motivations are as follows:
First of all, four space-time dimensions is the physical value
and it is the sunrise integral in four space-time dimensions which enters
precision calculations in high-energy particle physics.
Secondly, we hope to learn more about elliptic generalisations of polylogarithms.
In particular we are interested in generalisations with higher weight or depth.
Various elliptic generalisations of (multiple) polylogarithms have been discussed in the 
mathematical literature \cite{Beilinson:1994,Levin:1997,Levin:2007,Brown:2011,Wildeshaus,Bloch:2013tra,Adams:2014vja}.
The definitions of the generalised functions differ among the articles in the mathematical literature and depend on the postulated
mathematical properties one would like to keep in going from the (multiple) polylogarithms towards the elliptic case.
We are interested to find out, which definition nature has chosen in quantum field theory.

Unlike in two space-time dimensions, the sunrise integral in four space-time dimensions is divergent and requires regularisation.
Dimensional regularisation is the method of choice.
In $D=4-2\eps$ space-time dimensions, the $\eps$-expansion of the two-loop sunrise integral starts at $1/\eps^2$, corresponding to two
ultraviolet divergences.
The terms proportional to $1/\eps^2$ and $1/\eps$ are 
very well known and our main interest in this article are the terms proportional to $\eps^0$.

Using dimensional-shift relations we can relate the ${\mathcal O}(\eps^0)$-term around four space-time dimensions
to the ${\mathcal O}(\eps^0)$-term and the ${\mathcal O}(\eps^1)$-term
of the sunrise integral around two space-time dimensions.
In the equal mass case the coefficient of the ${\mathcal O}(\eps^1)$-term in this relation
vanishes and the ${\mathcal O}(\eps^0)$-term around four space-time dimensions
is related to the ${\mathcal O}(\eps^0)$-term around two space-time dimensions alone.
In this paper we treat the more general case of unequal masses and discuss as a specialisation the case of equal masses.

The ${\mathcal O}(\eps^0)$-term of the sunrise integral around two space-time dimensions
is already known \cite{Bloch:2013tra,Adams:2014vja}.
In this paper we compute the ${\mathcal O}(\eps^1)$-term of the sunrise integral around two space-time dimensions
and express it in terms of generalisations of Clausen and Glaisher functions towards the elliptic case.
We briefly recall that the Clausen and Glaisher functions are related to the real and imaginary parts of the classical polylogarithms.
Generalisations of multiple polylogarithms towards the elliptic case start 
to make their appearance in physics \cite{Bloch:2013tra,Adams:2014vja,Bloch:2014qca,Broedel:2014vla}.
The ${\mathcal O}(\eps^1)$-term of the sunrise integral around two space-time dimensions
gives us information on elliptic generalisations of multiple polylogarithms of depth greater than one.
In addition we observe that this term is not of uniform weight, it contains terms of weight three and four.
We discuss in detail the occurrence of the weight four terms.

This paper is organised as follows:
In section~\ref{sec:definition} we introduce the two-loop sunrise integral.
Section~\ref{sec:variables} is devoted to variables associated to this integral.
Our principal method to calculate the integral is based on differential equations, and the differential
equations obeyed by the two-loop sunrise integral are discussed in section~\ref{sec:dgl}.
The solution of any differential equation requires boundary values, and the boundary values for the
two-loop sunrise integral are presented in section~\ref{sec:boundary}.
In section~\ref{sec:Clausen} we introduce generalisations of the Clausen and Glaisher functions to the
elliptic setting. These functions will be needed to express our final result for the two-loop sunrise integral.
In section~\ref{sec:result_S1} we give the result for the ${\mathcal O}(\eps^1)$-part
of the sunrise integral around two space-time dimensions.
Using dimensional-shift identities 
we can relate the terms up to the finite part in the $\eps$-expansion of the two-loop sunrise integral 
around four space-time dimensions to the ${\mathcal O}(\eps^0)$-part and the ${\mathcal O}(\eps^1)$-part
of the sunrise integral around two space-time dimensions.
This is done in section~\ref{sec:shift}.
Section~\ref{sec:equal_mass_case} discusses the two-loop sunrise integral around
four space-time dimensions for the special case, where all masses are equal.
Section~\ref{sec:weights} is devoted to a detailed discussion of the transcendental weights occurring in our result
for the ${\mathcal O}(\eps^1)$-part
of the sunrise integral around two space-time dimensions.
Finally, section~\ref{sec:conclusions} contains our conclusions.
In an appendix we included explicit formulae, which are too long to be presented in the main text of this paper.

\section{Definition of the sunrise integral}
\label{sec:definition}

The two-loop integral corresponding to the sunrise graph with arbitrary masses is given 
in $D$-dimensional Minkowski space by
\bq
\label{def_sunrise}
\lefteqn{
 S_{\nu_1 \nu_2 \nu_3}\left( D, p^2, m_1^2, m_2^2, m_3^2, \mu^2 \right)
 = 
} & &
 \\
 & &
 \left(\mu^2\right)^{\nu-D}
 \int \frac{d^Dk_1}{i \pi^{\frac{D}{2}}} \frac{d^Dk_2}{i \pi^{\frac{D}{2}}}
 \frac{1}{\left(-k_1^2+m_1^2\right)^{\nu_1}\left(-k_2^2+m_2^2\right)^{\nu_2}\left(-\left(p-k_1-k_2\right)^2+m_3^2\right)^{\nu_3}}.
 \nonumber
\eq
In eq.~(\ref{def_sunrise}) the three internal masses are denoted by $m_1$, $m_2$ and $m_3$. 
Without loss of generality we assume that the masses are ordered as
\bq
\label{ordering_masses}
 0 < m_1 \le m_2 \le m_3.
\eq
The arbitrary scale $\mu$ is introduced to keep the integral dimensionless.
We denote by $\nu=\nu_1+\nu_2+\nu_3$ the sum of the exponents of the propagators.
The quantity $p^2$ denotes the momentum squared (with respect to the Minkowski metric)
and we will write
\bq
 t & = & p^2.
\eq
We perform our calculation in the region defined by
\bq
 -t & \ge & 0,
\eq
and in the vicinity of the equal mass point.
Our result can be continued analytically to all other regions of interest by
Feynman's $i0$-prescription, where we substitute $t \rightarrow t + i0$. 
The symbol $+i0$ denotes an infinitesimal positive imaginary part.

Where it is not essential we will suppress the dependence on the masses $m_i$ and the scale $\mu$ and simply write
$S_{\nu_1 \nu_2 \nu_3}( D, t)$ instead of $S_{\nu_1 \nu_2 \nu_3}( D, t, m_1^2, m_2^2, m_3^2, \mu^2)$.
In terms of Feynman parameters the two-loop integral is given by
\bq
\label{def_Feynman_integral}
 S_{\nu_1 \nu_2 \nu_3}\left( D, t\right)
 & = & 
 \frac{\Gamma\left(\nu-D\right)}{\Gamma\left(\nu_1\right)\Gamma\left(\nu_2\right)\Gamma\left(\nu_3\right)}
 \left(\mu^2\right)^{\nu-D}
 \int\limits_{\sigma} x_1^{\nu_1-1} x_2^{\nu_2-1} x_3^{\nu_3-1} \frac{{\cal U}^{\nu-\frac{3}{2}D}}{{\cal F}^{\nu-D}} \omega
\eq
with the two Feynman graph polynomials
\bq
 {\cal U} & = & x_1 x_2 + x_2 x_3 + x_3 x_1,
 \nonumber \\
 {\cal F} & = & - x_1 x_2 x_3 t
                + \left( x_1 m_1^2 + x_2 m_2^2 + x_3 m_3^2 \right) {\cal U}.
\eq
The differential two-form $\omega$ is given by
\bq
 \omega & = & x_1 dx_2 \wedge dx_3 - x_2 dx_1 \wedge dx_3 + x_3 dx_1 \wedge dx_2.
\eq
The integration is over
\bq
 \sigma & = & \left\{ \left[ x_1 : x_2 : x_3 \right] \in {\mathbb P}^2 | x_i \ge 0, i=1,2,3 \right\}.
\eq
In order to facilitate a comparison with results in the literature we remark that 
the definition of the sunrise integral in eq.~(\ref{def_sunrise}) is in Minkowski space. 
In a space with Euclidean signature one defines
the two-loop sunrise integral as
\bq
\lefteqn{
 S_{\nu_1 \nu_2 \nu_3}^{\mathrm{eucl}}\left( D, P^2, m_1^2, m_2^2, m_3^2, \mu^2 \right)
 = 
} & &
 \nonumber \\
 & &
 \left(\mu^2\right)^{\nu-D}
 \int \frac{d^DK_1}{\pi^{\frac{D}{2}}} \frac{d^DK_2}{\pi^{\frac{D}{2}}}
 \frac{1}{\left(K_1^2+m_1^2\right)^{\nu_1}\left(K_2^2+m_2^2\right)^{\nu_2}\left(\left(P-K_1-K_2\right)^2+m_3^2\right)^{\nu_3}}.
\eq
The momenta in Euclidean space are denoted by capital letters, while the ones in Minkowski space are denoted by lower case letters.
We have with $P^2=-p^2$ the relation
\bq
 S_{\nu_1 \nu_2 \nu_3}^{\mathrm{eucl}}\left( D, P^2, m_1^2, m_2^2, m_3^2, \mu^2 \right)
 & = &
 S_{\nu_1 \nu_2 \nu_3}\left( D, p^2, m_1^2, m_2^2, m_3^2, \mu^2 \right).
\eq
We are primarily concerned with the integral $S_{111}(4-2\eps,t)$.
This integral has a Laurent expansion in $\eps$, starting at $\eps^{-2}$:
\bq
\label{expansion_4D}
  S_{111}\left( 4-2\eps, t\right)
 & = &
 e^{-2 \gamma \eps} \left[ \frac{1}{\eps^2} S_{111}^{(-2)}(4,t) + \frac{1}{\eps} S_{111}^{(-1)}(4,t) + S_{111}^{(0)}(4,t)
 + {\cal O}\left(\eps\right) \right].
\eq
The first two terms $S_{111}^{(-2)}(4,t)$ and $S_{111}^{(-1)}(4,t)$ are rather simple, while the term $S_{111}^{(0)}(4,t)$
is in the focus of this paper.
With the help of dimensional-shift relations we relate the integral $S_{111}(4-2\eps,t)$ to the integral
$S_{111}(2-2\eps,t)$, which has an expansion in $\eps$ starting at $\eps^0$:
\bq
\label{expansion_2D}
  S_{111}\left( 2-2\eps, t\right)
 & = &
 e^{-2 \gamma \eps} \left[ S_{111}^{(0)}(2,t) + \eps S_{111}^{(1)}(2,t)
 + {\cal O}\left(\eps^2\right) \right].
\eq
The term $S_{111}^{(0)}(2,t)$ has been given in \cite{Adams:2014vja} in terms of elliptic dilogarithms,
the term $S_{111}^{(1)}(2,t)$ will be given in this paper.

In using dimensional-shift relations we will encounter simpler integrals, obtained by pinching internal propagators. 
For the sunrise integral any integral obtained by pinching propagators is a product of tadpole integrals.
The tadpole integral is given by
\bq
\label{def_tadpole}
 T_{\nu}\left( D, m^2, \mu^2 \right)
 & = &
 \left(\mu^2\right)^{\nu-\frac{D}{2}}
 \int \frac{d^Dk}{i \pi^{\frac{D}{2}}}
 \frac{1}{\left(-k^2+m^2\right)^{\nu}}
 \;\; = \;\;
 \frac{\Gamma\left(\nu-\frac{D}{2}\right)}{\Gamma\left(\nu\right)}
 \left( \frac{m^2}{\mu^2} \right)^{\frac{D}{2}-\nu}.
\eq

\section{Variables related to the sunrise integral}
\label{sec:variables}

The sunrise integral defined in eq.~(\ref{def_sunrise}) depends for a given space-time dimension $D$ 
on the variables $t$, $m_1^2$, $m_2^2$, $m_3^2$ and $\mu^2$.
It is clear from the definition that the sunrise integral will not change the value under a simultaneous
rescaling of all five quantities.
This implies that the integral depends only on the four dimensionless ratios $t/\mu^2$,
$m_1^2/\mu^2$, $m_2^2/\mu^2$ and $m_3^2/\mu^2$.
It will be convenient to view the non-trivial part of the integral as a function of five new variables
\bq
\label{set_new_variables}
 q, \;\; w_1, \;\; w_2, \;\; w_3 \;\; \mbox{and} \;\; \frac{m_1^2 m_2^2 m_3^2}{\mu^6}.
\eq
The variables $w_1$, $w_2$ and $w_3$ satisfy
\bq
 w_1 w_2 w_3 & = & 1,
\eq
therefore there are again only four independent variables.
The variables $q$, $w_1$, $w_2$ and $w_3$ are closely related 
to the elliptic curve defined by ${\mathcal F}=0$.

Before we discuss this change of variables we first introduce some convenient abbreviations.
Let us start by denoting the pseudo-thresholds by
\bq
\label{def_pseudo_thresholds}
 \mu_1 = m_1+m_2-m_3,
 \;\;\;
 \mu_2 = m_1-m_2+m_3,
 \;\;\;
 \mu_3 = -m_1+m_2+m_3,
\eq
and the threshold by
\bq
\label{def_thresholds}
 \mu_4 = m_1+m_2+m_3.
\eq
In the vicinity of the equal mass point we can assume $m_3 < m_1+m_2$, or equivalently $\mu_1>0$.
With this assumption and with the ordering of eq.~(\ref{ordering_masses}) we have
\bq
 0 < \mu_1 \le \mu_2 \le \mu_3 < \mu_4.
\eq
We further set
\bq
\label{def_D}
 D & = & 
 \left( t - \mu_1^2 \right)
 \left( t - \mu_2^2 \right)
 \left( t - \mu_3^2 \right)
 \left( t - \mu_4^2 \right).
\eq
The variable $D$ will appear in the algebraic prefactor of our result.
It will be convenient to introduce 
the monomial symmetric polynomials $M_{\lambda_1 \lambda_2 \lambda_3}$ in the variables $m_1^2$, $m_2^2$ and $m_3^2$.
These are defined by
\bq
 M_{\lambda_1 \lambda_2 \lambda_3} & = &
 \sum\limits_{\sigma} \left( m_1^2 \right)^{\sigma\left(\lambda_1\right)} \left( m_2^2 \right)^{\sigma\left(\lambda_2\right)} \left( m_3^2 \right)^{\sigma\left(\lambda_3\right)},
\eq
where the sum is over all distinct permutations of $\left(\lambda_1,\lambda_2,\lambda_3\right)$. A few examples are
\bq
 M_{100} 
 & = & 
 m_1^2 + m_2^2 + m_3^2,
 \nonumber \\
 M_{111} 
 & = & 
 m_1^2 m_2^2 m_3^2,
 \nonumber \\
 M_{210} 
 & = & 
 m_1^4 m_2^2 + m_2^4 m_3^2 + m_3^4 m_1^2 + m_2^4 m_1^2 + m_3^4 m_2^2 + m_1^4 m_3^2.
\eq
In addition, we introduce the abbreviations
\bq
\label{def_delta}
\Delta = \mu_1 \mu_2 \mu_3 \mu_4,
 \;\;\;\;\;\;
 \delta_1 = -m_1^2 + m_2^2 + m_3^2,
 \;\;\;\;\;\;
 \delta_2 = m_1^2 - m_2^2 + m_3^2,
 \;\;\;\;\;\;
 \delta_3 = m_1^2 + m_2^2 - m_3^2
\eq
and
\bq
\label{def_v_variables}
 v_1 = \frac{1+i\frac{\sqrt{\Delta}}{\delta_1}}{1-i\frac{\sqrt{\Delta}}{\delta_1}},
 \;\;\;\;\;\;
 v_2 = \frac{1+i\frac{\sqrt{\Delta}}{\delta_2}}{1-i\frac{\sqrt{\Delta}}{\delta_2}},
 \;\;\;\;\;\;
 v_3 = \frac{1+i\frac{\sqrt{\Delta}}{\delta_3}}{1-i\frac{\sqrt{\Delta}}{\delta_3}}.
\eq
In the vicinity of the equal mass point we have
\bq
 \Delta > 0,
 \;\;\;\;\;\;
 \delta_i > 0, & & i \in \{1,2,3\}.
\eq
The variables $v_1$, $v_2$ and $v_3$ are 
then
complex numbers of unit norm.
In addition they satisfy
\bq
 v_1 v_2 v_3 & = & 1.
\eq
We now describe in detail the change of variables from the set $t, m_1^2, m_2^2, m_3^2, \mu^2$ to the set specified in eq.~(\ref{set_new_variables}).
The equation
\bq
 {\mathcal F} & = & 0
\eq
is a polynomial equation of degree three in the Feynman parameters $x_1$, $x_2$ and $x_3$.
The cubic curve intersects the domain of integration in the three points
\bq
\label{intersection_F_sigma}
 P_1 = \left[1:0:0\right], 
 \;\;\;
 P_2 = \left[0:1:0\right], 
 \;\;\;
 P_3 = \left[0:0:1\right].
\eq
The cubic curve together with a choice of a point $O \in \{ P_1, P_2, P_3 \}$ as origin defines an elliptic curve.
All three choices will lead to the same Weierstrass normal form, given by
\bq
 y^2 
 & = & 
 4 \left(x-e_1\right)\left(x-e_2\right)\left(x-e_3\right),
 \;\;\;\;\;\;
 \mbox{with} 
 \;\;\;
 e_1+e_2+e_3=0,
\eq
and
\bq
\label{def_roots}
 e_1 
 & = & 
 \frac{1}{24 \mu^4} \left( -t^2 + 2 M_{100} t + \Delta + 3 \sqrt{D} \right),
 \nonumber \\
 e_2 
 & = & 
 \frac{1}{24 \mu^4} \left( -t^2 + 2 M_{100} t + \Delta - 3 \sqrt{D} \right),
 \nonumber \\
 e_3 
 & = & 
 \frac{1}{24 \mu^4} \left( 2 t^2 - 4 M_{100} t - 2 \Delta \right).
\eq
The modulus $k$ and the complementary modulus $k'$ of the elliptic curve are given by
\bq
\label{def_modulus}
 k = \sqrt{\frac{e_3-e_2}{e_1-e_2}},
 & &
 k' = \sqrt{1-k^2} = \sqrt{\frac{e_1-e_3}{e_1-e_2}}.
\eq
The periods of the elliptic curve can be taken as
\bq
\label{def_periods}
 \psi_1 =  
 2 \int\limits_{e_2}^{e_3} \frac{dx}{y}
 =
 \frac{4 \mu^2}{D^{\frac{1}{4}}} K\left(k\right),
 & &
 \psi_2 =  
 2 \int\limits_{e_1}^{e_3} \frac{dx}{y}
 =
 \frac{4 i \mu^2}{D^{\frac{1}{4}}} K\left(k'\right),
 \\
 \phi_1 =  
 \frac{8\mu^4}{D^{\frac{1}{2}}} \int\limits_{e_2}^{e_3} \frac{\left(x-e_2\right) dx}{y}
 =
 \frac{4 \mu^2}{D^{\frac{1}{4}}} \left( K\left(k\right)- E\left(k\right) \right),
 & &
 \phi_2 =  
 \frac{8\mu^4}{D^{\frac{1}{2}}} \int\limits_{e_1}^{e_3} \frac{\left(x-e_2\right) dx}{y}
 =
 \frac{4 i \mu^2}{D^{\frac{1}{4}}} E\left(k'\right).
 \nonumber
\eq
$K(x)$ and $E(x)$ denote the complete elliptic integral of the first kind and second kind, respectively:
\bq
 K(x)
 = 
 \int\limits_0^1 \frac{dt}{\sqrt{\left(1-t^2\right)\left(1-x^2t^2\right)}},
 & &
 E(x)
 = 
 \int\limits_0^1 dt \sqrt{\frac{1-x^2t^2}{1-t^2}}.
\eq
The Legendre relation for the periods reads
\bq
 \psi_1 \phi_2 - \psi_2 \phi_1
 & = &
 \frac{8 \pi i \mu^4}{D^{\frac{1}{2}}}.
\eq
The Wronskian is given by
\bq
\label{def_Wronski}
 W & = &
 \psi_1 \frac{d}{dt} \psi_2 - \psi_2 \frac{d}{dt} \psi_1
 =
 -
 4 \pi i \mu^4 
 \;
 \frac{\left( 3 t^2 - 2 M_{100} t + \Delta \right)}{t\left( t - \mu_1^2 \right)\left( t - \mu_2^2 \right)\left( t - \mu_3^2 \right)\left( t - \mu_4^2 \right)}.
\eq
We denote the ratio of the two periods $\psi_2$ and $\psi_1$ by
\bq
 \tau 
 & = & 
 \frac{\psi_2}{\psi_1}
\eq
and the nome by
\bq
\label{def_nome}
 q & = & e^{i\pi \tau}.
\eq
The nome $q$ will be one of our new variables, and we can think of the variable $q$ as replacing
the variable $t$.
A useful relation for expressing $q$ in terms of $t$ (or vice versa) is given by
\bq
\label{basic_relation_power_series}
 \frac{t}{\left( \mu_1^2 - t \right) \left( \mu_2^2 - t \right) \left( \mu_3^2 - t\right) \left( \mu_4^2 -t \right)}
 & = &
 - \frac{1}{m_1^2 m_2^2 m_3^2}
 \frac{\eta\left(\frac{\tau}{2}\right)^{24}\eta\left(2\tau\right)^{24}}{\eta\left(\tau\right)^{48}}.
\eq
In this equation, Dedekind's $\eta$-function, defined by
\bq
 \eta\left(\tau\right)
 & = &
 e^{\frac{\pi i \tau}{12}} \prod\limits_{n=1}^\infty \left( 1- e^{2 \pi i n \tau} \right)
 =
 q^{\frac{1}{12}} \prod\limits_{n=1}^\infty \left( 1 - q^{2n} \right),
\eq
appears. The left-hand side of eq.~(\ref{basic_relation_power_series}) 
has a power series in $t$, the right-hand side
of eq.~(\ref{basic_relation_power_series}) has a power series in $q$.
Eq.~(\ref{basic_relation_power_series}) can be used to express $t$ as a power series in $q$,
or vice versa.
We have
\bq
 t = 
 - q \frac{\Delta^2}{m_1^2 m_2^2 m_3^2}
 + {\cal O}\left(q^2\right),
 & &
 q = 
 - t \frac{m_1^2 m_2^2 m_3^2}{\Delta^2}
 + {\cal O}\left(t^2\right),
\eq
therefore $t \rightarrow 0$ implies $q \rightarrow 0$ and vice versa.
It remains to define the variables $w_1$, $w_2$ and $w_3$. These are given by
\bq
\label{def_arguments_w_i}
 w_i 
 = e^{i \beta_i},
 \;\;\;\;
 \beta_i 
 = \pi \frac{F\left(u_i,k\right)}{K\left(k\right)},
 \;\;\;\;
 u_i 
 = \sqrt{\frac{e_1-e_2}{x_{j,k}-e_2}},
 \;\;\;\;
 x_{j,k} 
 = e_3 + \frac{m_j^2 m_k^2}{\mu^4}.
\eq
In the definition of $u_i$ we used the convention that $(i,j,k)$ is a permutation of $(1,2,3)$.
In the definition of $\beta_i$ the incomplete elliptic integral of the first kind appears, 
defined by
\bq
 F\left(z,x\right)
 & = &
 \int\limits_0^z \frac{dt}{\sqrt{\left(1-t^2\right)\left(1-x^2t^2\right)}}.
\eq
In the case $q=0$ (or equivalently $t=0$) we have
\bq
 \lim\limits_{q \rightarrow 0} w_j & = & v_j,
 \;\;\;\;\;\;\;\;\; j \in \{ 1,2,3 \},
\eq
while in the equal mass case $m_1=m_2=m_3$ we have
\bq
 \left. w_j \right|_{m_1=m_2=m_3}
 & = & e^{\frac{2\pi i}{3}},
 \;\;\;\;\;\;\;\;\; j \in \{ 1,2,3 \}
\eq
for all $q$ (or equivalently all $t$).
There is a simple geometric interpretation for the variables $w_1$, $w_2$ and $w_3$:
We recall that the points $P_1$, $P_2$ and $P_3$
are the intersection points of ${\mathcal F}=0$ with the integration region $\sigma$.
One of these points is chosen as origin of the elliptic curve.
The set
\bq
 \left\{ w_1, w_2, w_3, w_1^{-1}, w_2^{-1}, w_3^{-1} \right\}
\eq
is obtained from the images in the Jacobi uniformisation of the two points not chosen as origin
by considering all three choices of origins.
More details can be found in \cite{Adams:2014vja}.

\section{Differential equations}
\label{sec:dgl}

In this section we present the (fourth-order) differential equation for the sunrise integral $S_{111}(D,t)$ in $D$ space-time dimensions 
with arbitrary masses.
From this equation we deduce a differential equation for the ${\mathcal O}(\eps^1)$-piece $S_{111}^{(1)}(2,t)$.
The latter differential equation is then solved up to quadrature.

In two space-time dimensions the integral $S_{111}(2,t)=S_{111}^{(0)}(2,t)$ satisfies 
a second-order differential equation \cite{MullerStach:2011ru}:
\bq
\label{second_order_dgl}
 \left[ p_2 \frac{d^2}{d t^2} + p_1 \frac{d}{dt} + p_0 \right] S_{111}^{(0)}\left(2,t\right) & = & \mu^2 p_3,
\eq
The coefficients $p_2$, $p_1$ and $p_0$ as well as $p_3$ are collected in appendix~\ref{appendix:coeff}.
The left hand side defines a Picard-Fuchs operator 
\bq
\label{def_L2}
 L^{(0)}_2\left(2\right) & = &
 p_2 \frac{d^2}{d t^2} + p_1 \frac{d}{dt} + p_0. 
\eq
The periods $\psi_1$ and $\psi_2$ are solutions of the homogeneous differential equation \cite{Adams:2013nia}
\bq
 L^{(0)}_2\left(2\right) \psi_i & = & 0,
 \;\;\;\;\;\;\;\;\;
 i \in \{ 1,2 \}.
\eq
We will encounter several differential operators.
We will use the notation
\bq
 L^{(j)}_{r,i}\left(D\right),
\eq
where $r$ denotes the order of the differential operator and $D$ denotes the associated space-time dimension.
In the case where $D$ is an integer and a superscript $j$ is present, this superscript
denotes the order in the $\eps$-expansion to which this operator belongs.
Finally, $i$ is a label to distinguish differential operators with identical $r$, $D$, $j$.

In $D$ dimensions the integral $S_{111}(D,t)$ satisfies a fourth-order differential equation.
\bq
\label{diff_eq_D_dim}
 \left[
 P_4 \frac{d^4}{dt^4}
 + 
 P_3 \frac{d^3}{dt^3}
 + 
 P_2 \frac{d^2}{dt^2}
 + 
 P_1 \frac{d}{dt}
 + 
 P_0
 \right] S_{111}\left(D,t\right)
 & = &
 \mu^2
 \left[
 c_{12} T_{12}
 +
 c_{13} T_{13}
 +
 c_{23} T_{23}
 \right].
\eq
Here we used the abbreviation 
\bq
 T_{ij} & = & 
 T_1\left(D,m_i^2,\mu^2\right) T_1\left(D,m_j^2,\mu^2\right).
\eq
The explicit expressions for the coefficients $P_4$, $P_3$, $P_2$, $P_1$, $P_0$ and $c_{12}$, $c_{13}$, $c_{23}$ are rather long and given
in appendix~\ref{appendix:coeff}.
There are several possibilities to obtain the fourth-order differential equation:
It can either be derived by using the relations given in \cite{Caffo:1998du}, 
by using the program ``Reduze'' \cite{Studerus:2009ye,vonManteuffel:2012np} or by the algorithm given in \cite{MullerStach:2012mp}.
Eq.~(\ref{diff_eq_D_dim}) defines a fourth-order Picard-Fuchs operator
\bq
 L_4\left(D\right) 
 & = & 
 P_4 \frac{d^4}{dt^4}
 + 
 P_3 \frac{d^3}{dt^3}
 + 
 P_2 \frac{d^2}{dt^2}
 + 
 P_1 \frac{d}{dt}
 + 
 P_0.
\eq
The Picard-Fuchs operator $L_4(D)$ has a polynomial
dependence on the number of space-time dimensions $D$.
Around $D=2-2\eps$ we can write
\bq
 L_4\left(2-2\eps\right)
 & = &
 \sum\limits_{j=0}^5 \eps^j \; L^{(j)}_4\left(2\right).
\eq
Of particular relevance for the ${\mathcal O}(\eps^1)$-part $S_{111}^{(1)}(2,t)$
will be the operators $L^{(0)}_4(2)$ and $L^{(1)}_4(2)$.
The operator $L^{(0)}_4(2)$ factorises
\bq
 L^{(0)}_4\left(2\right) & = & 
 L^{(0)}_{1,a}\left(2\right)
 \;\; 
 L^{(0)}_{1,b}\left(2\right)
 \;\;
 L^{(0)}_{2}\left(2\right).
\eq
The differential operator $L^{(0)}_{2}(2)$ is the one we already encountered in eq.~(\ref{def_L2}).
The two other factors $L^{(0)}_{1,a}(2)$ and $L^{(0)}_{1,b}(2)$ are first-order differential operators:
\bq
 L^{(0)}_{1,a}\left(2\right) 
 & = &
 8 t^2
 \left[ 
 \frac{\left(5t^2 + 2M_{100} t + 7 \Delta \right)}{\left( 15 t^2 - 2 M_{100} t - 3 \Delta \right) \left(3t^2-2 M_{100}t+\Delta\right)} \frac{d}{dt}
 \right. \nonumber \\
 & & \left.
 - \frac{60 t^3 - 12 M_{100} t^2 - \left(60 M_{200} - 88 M_{110} \right) t - 12 M_{100} \Delta}{\left( 15 t^2 - 2 M_{100} t - 3 \Delta \right) \left(3t^2-2 M_{100}t+\Delta\right)^2}
 \right],
 \nonumber \\
 L^{(0)}_{1,b}\left(2\right) 
 & = &
 \left( 15 t^2 - 2 M_{100} t - 3 \Delta \right) \frac{d}{dt} - \left( 30 t - 2 M_{100} \right).
\eq
Solutions to the homogeneous equations
\bq
 L^{(0)}_{1,a}\left(2\right) \; \psi_{a}\left(t\right) = 0,
 & &
 L^{(0)}_{1,b}\left(2\right) \; \psi_{b}\left(t\right) = 0
\eq
are
\bq
 \psi_{a}\left(t\right) & = &
 \left( 3 t^2 - 2 M_{100} t + \Delta \right)
 \left( 5 t^2 + 2 M_{100} t + 7 \Delta \right),
 \nonumber \\
 \psi_{b}\left(t\right) & = & 15 t^2 - 2 M_{100} t - 3 \Delta.
\eq
Substituting the $\eps$-expansion of $S_{111}(2-2\eps,t)$ given in eq.~(\ref{expansion_2D}) 
into the $D$-dimensional differential equation~(\ref{diff_eq_D_dim}) gives a coupled system of differential
equations for $S_{111}^{(j)}(2,t)$, where the differential equation for $S_{111}^{(j)}(2,t)$ will involve
the lower order integrals $S_{111}^{(i)}(2,t)$ with $i<j$. 
This system can be solved order by order in $\eps$.
At 
order $\eps^0$ 
one finds 
\bq
\label{dgl_S_111_0}
 L^{(0)}_{1,a}\left(2\right)
 \;\; 
 L^{(0)}_{1,b}\left(2\right) 
 \;\;
 L^{(0)}_{2}\left(2\right)
 \;\;
 S_{111}^{(0)}(2,t)
 & = &
 - 32 \mu^2 t^2 \left( 15 t^2 + 14 M_{100} t + 77 \Delta \right).
\eq
From eq.~(\ref{second_order_dgl}) we know already that
\bq
 L^{(0)}_{2}\left(2\right)
 \;\;
 S_{111}^{(0)}(2,t)
 & = &
 \mu^2 p_3(t),
\eq
and eq.~(\ref{dgl_S_111_0}) reduces to
\bq
 \mu^2
 \;\;
 L^{(0)}_{1,a}\left(2\right)
 \;\; 
 L^{(0)}_{1,b}\left(2\right) 
 \;\;
 p_3(t)
 & = &
 - 32 \mu^2 t^2 \left( 15 t^2 + 14 M_{100} t + 77 \Delta \right),
\eq
which is easily verified.
\\
\\
At order $\eps^1$ we have
\bq
\label{dgl_S_111_1}
 L^{(0)}_{1,a}\left(2\right)
 \;\; 
 L^{(0)}_{1,b}\left(2\right) 
 \;\;
 L^{(0)}_{2}\left(2\right)
 \;\;
 S_{111}^{(1)}(2,t)
 & = &
 I_1\left(t\right),
\eq
with
\bq
\label{def_I_t}
 I_1\left(t\right)
 & = &
 - L_4^{(1)}(2) \;\; S_{111}^{(0)}(2,t)
 \nonumber \\
 & &
 - \mu^2 
 \left\{
  912 t^4 + 1344 M_{100} t^3 + \left( 9088 M_{110} - 4416 M_{200} \right) t^2
  + 512 M_{100} \Delta t
  + 112 \Delta^2
 \right. \nonumber \\
 & & \left.
  + d\left(t,m_1^2,m_2^2,m_3^2\right) \ln \frac{m_1^2}{\mu^2}
  + d\left(t,m_2^2,m_3^2,m_1^2\right) \ln \frac{m_2^2}{\mu^2}
  + d\left(t,m_3^2,m_1^2,m_2^2\right) \ln \frac{m_3^2}{\mu^2}
 \right\}
\eq
and
\bq
\lefteqn{
 d\left(t,m_1^2,m_2^2,m_3^2\right)
 = 
 -320 t^4
 - \left( 352 m_1^2 + 272 m_2^2 + 272 m_3^2 \right) t^3
 } & & \nonumber \\
 & &
 + \left( 1440 m_1^4 + 1744 m_2^4 + 1744 m_3^4 - 3120 m_1^2 m_2^2 - 3120 m_1^2 m_3^2 - 3616 m_2^2 m_3^2 \right) t^2
 \nonumber \\
 & &
 + \left( -544 m_1^6 + 272 m_2^6 + 272 m_3^6 + 1296 m_1^4 m_2^2 + 1296 m_1^4 m_3^2 - 1024 m_1^2 m_2^4 - 1024 m_1^2 m_3^4 
 \right. \nonumber \\
 & & \left. - 272 m_2^4 m_3^2 - 272 m_2^2 m_3^4 \right) t
 \nonumber \\
 & &
 + \left( 224 m_1^4 - 112 m_2^4 - 112 m_3^4 - 112 m_1^2 m_2^2 - 112 m_1^2 m_3^2 + 224 m_2^2 m_3^2 \right) \Delta.
\eq
Eq.~(\ref{dgl_S_111_1}) is a fourth-order differential equation for the integral $S_{111}^{(1)}(2,t)$.
The fourth-order differential operator in eq.~(\ref{dgl_S_111_1}) factorises 
into two first-order differential operators
and a second-order differential operator.
The inhomogeneous term $I_1(t)$ has a Taylor expansion in $t$, starting with $t^2$.
Eq.~(\ref{dgl_S_111_1}) is easily solved for $L^{(0)}_{2}(2) \; S_{111}^{(1)}(2,t)$:
\bq
\label{L2_S_111_1}
 L^{(0)}_{2}\left(2\right) \;\; S_{111}^{(1)}(2,t)
 & = & 
 I_2\left(t\right),
\eq
with
\bq
\lefteqn{
 I_2\left(t\right)
 = }
 & & \\
 & &
 C_1 \psi_b(t) 
 + C_2 \psi_b(t) \int\limits_0^t \frac{\psi_a(t_1) dt_1}{p_{1,b}(t_1) \psi_b(t_1)}  
 + \psi_b(t) \int\limits_0^t \frac{\psi_a(t_1) dt_1}{p_{1,b}(t_1) \psi_b(t_1)} 
             \int\limits_0^{t_1} \frac{I_1\left(t_2\right) dt_2}{p_{1,a}(t_2) \psi_a(t_2)},
 \nonumber
\eq
where $C_1$ and $C_2$ are two integration constants to be determined from the boundary conditions.
In eq.~(\ref{L2_S_111_1}) we used the notation that
\bq
 L^{(0)}_{1,a}\left(2\right)
 =  
 p_{1,a} \frac{d}{dt} + p_{0,a},
 & &
 L^{(0)}_{1,b}\left(2\right)
 =  
 p_{1,b} \frac{d}{dt} + p_{0,b}.
\eq
For the homogeneous solutions we have
\bq
 C_1 \psi_b(t) & = &
 C_1 \left( 15 t^2 - 2 M_{100} t - 3 \Delta \right),
 \nonumber \\
 C_2 \psi_b(t) \int\limits_0^t \frac{\psi_a(t_1) dt_1}{p_{1,b}(t_1) \psi_b(t_1)} 
 & = &
 \frac{C_2}{3} \left(3 t^2 + 6 M_{100} t - 7 \Delta \right) t.
\eq
The differential equation~(\ref{L2_S_111_1}) is now of the same type as eq.~(\ref{second_order_dgl}),
only the inhomogeneous term differs.
It can be solved with the same methods as used for eq.~(\ref{second_order_dgl}) by changing variables from $t$ 
to the nome $q$. This change of variables is described in detail in \cite{Adams:2014vja}.
One finds
\bq
\label{solution_S1_quadrature}
 S_{111}^{(1)}(2,t)
 & = &
 C_3 \psi_1 + C_4 \psi_2
 - 
 \frac{\psi_1}{\pi}
 \int\limits_0^q \frac{dq_1}{q_1}
 \int\limits_0^{q_1} \frac{dq_2}{q_2}
 \;
 \frac{I_2\left(q_2\right) \psi_1\left(q_2\right)^3}{\pi p_2\left(q_2\right) W\left(q_2\right)^2}.
\eq
$C_3$ and $C_4$ are two further integration constants, which need to be determined from boundary conditions.
Eq.~(\ref{solution_S1_quadrature}) gives the solution for the integral $S_{111}^{(1)}(2,t)$ up to quadrature.
We still need to determine the integration constants. In addition we would like 
to express the integral $S_{111}^{(1)}(2,t)$ in terms of elliptic generalisations of multiple polylogarithms.
Let us also mention that eq.~(\ref{solution_S1_quadrature}) can be used to obtain the $q$-expansion (or equivalently the $t$-expansion) of
$S_{111}^{(1)}(2,t)$ to high orders in $q$ (or $t$).

\section{Boundary values}
\label{sec:boundary}

In this section we give the boundary values at $t=0$ for the first two coefficients of the $\eps$-expansion of $S_{111}(2-2\eps,t)$
(i.e. $S_{111}^{(0)}(2,0)$ and $S_{111}^{(1)}(2,0)$)
and for the first three coefficients of the $\eps$-expansion of $S_{111}(4-2\eps,t)$
(i.e. $S_{111}^{(-2)}(4,0)$, $S_{111}^{(-1)}(4,0)$ and $S_{111}^{(0)}(4,0)$)
We will use the boundary values $S_{111}^{(1)}(2,0)$ and $S_{111}^{(0)}(4,0)$ together with regularity conditions at $t=0$ to fix the integration
constants $C_1$, $C_2$, $C_3$ and $C_4$ of the previous section.

Let us start with the boundary value at $t=0$ of the sunrise integral in $D=2-2\eps$ dimensions.
We have
\bq
 S_{111}\left( 2-2\eps, 0\right)
 & = &
 \Gamma\left(1+2\eps\right)
 \left(\mu^2\right)^{1+2\eps}
 \int\limits_{\sigma} \frac{\omega}{\left( x_1 m_1^2 + x_2 m_2^2 + x_3 m_3^2 \right)^{1+2\eps} {\cal U}^{1-\eps}}.
\eq
By a change of variables we can relate this integral 
to the one-loop three-point function 
in $4+2\eps$ space-time dimensions (please note the sign of the $\eps$-part)
with massless internal lines
and three external masses.
The change of variables can be found in \cite{Adams:2013nia} and the result of the one-loop three point function can be taken from
\cite{Bern:1994kr,Lu:1992ny}.
One obtains for the $\eps$-expansion of $S_{111}( 2-2\eps, 0)$
\bq
 \sum\limits_{j=0}^\infty \eps^j S_{111}^{(j)}\left(2,0\right)
 & = &
 e^{2 \gamma \eps} \Gamma\left(1+2\eps\right)
 \left( \frac{\sqrt{\Delta}}{\mu^2} \right)^{-1-2\eps}
 \left[ \frac{1}{2\eps^2} \frac{\Gamma\left(1+\eps\right)^2}{\Gamma\left(1+2\eps\right)}
        \left( f_1+f_2+f_3\right)
        - \frac{\pi}{\eps} \right],
\eq
with
\bq
 f_j
 & = &
 \frac{1}{i}
 \left[ 
  \left(-v_j\right)^{-\eps} \; {}_2F_1\left(-2\eps,-\eps;1-\eps; v_j \right)
  -
  \left(-v_j^{-1}\right)^{-\eps} \; {}_2F_1\left(-2\eps,-\eps;1-\eps; v_j^{-1} \right)
 \right].
\eq
The expansion of the hypergeometric function reads
\bq
 {}_2F_1\left(-2\eps,-\eps;1-\eps; x \right)
 & = &
 1 + 2 \eps^2 \mathrm{Li}_2\left(x\right)
 + \eps^3 \left[ 2 \mathrm{Li}_3\left(x\right) - 4 \mathrm{Li}_{2,1}\left(x,1\right) \right]
 + {\mathcal O}\left(\eps^4\right).
\eq
We obtain for $S_{111}^{(0)}(2,0)$ and $S_{111}^{(1)}(2,0)$
\bq
 S_{111}^{(0)}\left(2,0\right)
 & = &
 \frac{2\mu^2}{\sqrt{\Delta}}
 \sum\limits_{j=1}^3
 \frac{1}{2i} 
 \left[ \mathrm{Li}_2\left(v_j\right) - \mathrm{Li}_2\left(v_j^{-1}\right) \right],
 \nonumber \\
 S_{111}^{(1)}\left(2,0\right)
 & = &
 \frac{2\mu^2}{\sqrt{\Delta}}
 \sum\limits_{j=1}^3
 \frac{1}{2i} 
 \left\{
  - 2 \mathrm{Li}_{2,1}\left(v_j,1\right) - \mathrm{Li}_3\left(v_j\right) 
  + 2  \mathrm{Li}_{2,1}\left(v_j^{-1},1\right) + \mathrm{Li}_3\left(v_j^{-1}\right)
 \right. \nonumber \\
 & & \left.
  - 2 \ln\left(\frac{\sqrt{\Delta}}{\mu^2}\right) \left[ \mathrm{Li}_2\left(v_j\right) - \mathrm{Li}_2\left(v_j^{-1}\right) \right]
 \right\}.
\eq
The definition of the multiple polylogarithms is given in eq.~(\ref{def_multiple_polylogs}).
Let us now turn to the values of the coefficients of the Laurent expansion of $S_{111}(4-2\eps,t)$ at $t=0$.
At $t=0$ we find 
\bq
\label{boundary_t_0}
\lefteqn{
 S_{111}^{(-2)}(4,0)
 =
 - \frac{M_{100}}{2 \mu^2},
} & &
 \nonumber \\
\lefteqn{
 S_{111}^{(-1)}(4,0)
 =
          - \frac{3 M_{100}}{2 \mu^2} 
          + \frac{m_1^2}{\mu^2} \ln\left(\frac{m_1^2}{\mu^2}\right)
          + \frac{m_2^2}{\mu^2} \ln\left(\frac{m_2^2}{\mu^2}\right)
          + \frac{m_3^2}{\mu^2} \ln\left(\frac{m_3^2}{\mu^2}\right),
} & &
 \nonumber \\
\lefteqn{
 S_{111}^{(0)}(4,0)
 =
 \frac{\Delta}{2\mu^4} S_{111}^{(0)}(2,0)
  - \frac{7 M_{100}}{2\mu^2}
  + 3 
      \left[ \frac{m_1^2}{\mu^2} \ln\left(\frac{m_1^2}{\mu^2}\right)
           + \frac{m_2^2}{\mu^2} \ln\left(\frac{m_2^2}{\mu^2}\right)
           + \frac{m_3^2}{\mu^2} \ln\left(\frac{m_3^2}{\mu^2}\right)
      \right]
} & &
 \nonumber \\
 & &
  - \frac{1}{2}
      \left[ \frac{m_1^2}{\mu^2} \ln^2\left(\frac{m_1^2}{\mu^2}\right)
           + \frac{m_2^2}{\mu^2} \ln^2\left(\frac{m_2^2}{\mu^2}\right)
           + \frac{m_3^2}{\mu^2} \ln^2\left(\frac{m_3^2}{\mu^2}\right)
      \right]
 \nonumber \\
 & &
  - \frac{1}{2\mu^2}
   \left[ 
          \left( m_1^2 + m_2^2 - m_3^2 \right) \ln\left(\frac{m_1^2}{\mu^2}\right) \ln\left(\frac{m_2^2}{\mu^2}\right)
        + \left( m_1^2 - m_2^2 + m_3^2 \right) \ln\left(\frac{m_1^2}{\mu^2}\right) \ln\left(\frac{m_3^2}{\mu^2}\right)
 \right. \nonumber \\
 & & \left.
        + \left( -m_1^2 + m_2^2 + m_3^2 \right) \ln\left(\frac{m_2^2}{\mu^2}\right) \ln\left(\frac{m_3^2}{\mu^2}\right)
   \right]
 - \frac{M_{100}}{2 \mu^2} \zeta_2.
\eq
The expression for $S_{111}^{(0)}(4,0)$ is obtained from the expression given in \cite{Caffo:1998du} 
by noting that the function
\bq
 L & = &
 \mathrm{Li}_2\left(- \frac{m_2}{m_1} t_3 \right)
 + \mathrm{Li}_2\left(- \frac{m_1}{m_2} t_3 \right)
 + \zeta_2
 + \frac{1}{2}\ln^2 t_3
 \nonumber \\
 & &
 + \frac{1}{2} \left[
         \ln\left(t_3+\frac{m_2}{m_1}\right)
        - \ln\left(t_3+\frac{m_1}{m_2}\right)
        + \frac{3}{4} \ln\frac{m_1^2}{m_2^2}
   \right]
   \ln\frac{m_1^2}{m_2^2},
 \nonumber \\
 t_3 & = &
 \frac{1}{2m_1m_2} \left( m_3^2 - m_1^2 - m_2^2 + \sqrt{-\Delta} \right)
\eq
can be written in a more symmetrical way as 
\bq
 L & = &
 - \frac{1}{2}
 \left[
        \mathrm{Li}_2\left( v_1 \right) + \mathrm{Li}_2\left( v_2 \right) + \mathrm{Li}_2\left( v_3 \right) 
      - \mathrm{Li}_2\left( v_1^{-1} \right) - \mathrm{Li}_2\left( v_2^{-1} \right) - \mathrm{Li}_2\left( v_3^{-1} \right) 
 \right].
\eq
The constants of integration $C_1$, $C_2$, $C_3$ and $C_4$ are determined as follows:
In section~\ref{sec:shift} we discuss dimensional-shift relations.
These relations allow us to relate the integral $S_{111}(4-2\eps,t)$ to the integral $S_{111}(2-2\eps,t)$, and we can determine
the integration constants from the following four conditions:
\begin{enumerate}
\item The requirement that $S_{111}^{(0)}(4,t)$ is regular at $t=0$, i.e. there is no pole at $t=0$ for $S_{111}^{(0)}(4,t)$.
\item The boundary value $S_{111}^{(0)}(4,0)$.
\item The requirement that $S_{111}^{(1)}(2,t)$ is regular at $t=0$, i.e. there is no logarithmic singularity at $t=0$ for $S_{111}^{(1)}(2,t)$.
\item The boundary value $S_{111}^{(1)}(2,0)$.
\end{enumerate}
The explicit expressions for $C_1$ and $C_2$ are given in appendix~\ref{section_integration_constants}.
Condition $3$ implies $C_4=0$.
With
\bq
 \psi_1\left(t=0\right) & = & \frac{2 \pi \mu^2}{\sqrt{\Delta}}
\eq
one finds for $C_3$
\bq
 C_3 
 & = &
 \frac{\sqrt{\Delta}}{2\pi \mu^2} S_{111}^{(1)}\left(2,0\right).
\eq

\section{Generalisations of the Clausen and Glaisher functions}
\label{sec:Clausen}

In this section we introduce elliptic generalisations of the Clausen and Glaisher functions.
These generalisations will show up in our final result.

The classical polylogarithms are defined by
\bq
 \mathrm{Li}_n\left(x\right) & = & \sum\limits_{j=1}^\infty \; \frac{x^j}{j^n},
\eq
and the multiple polylogarithms by
\bq
\label{def_multiple_polylogs}
 \mathrm{Li}_{n_1,n_2,...,n_k}\left(x_1,x_2,...,x_k\right)
 & = &
 \sum\limits_{j_1=1}^\infty \sum\limits_{j_2=1}^{j_1-1} ... \sum\limits_{j_k=1}^{j_{k-1}-1}
 \frac{x_1^{j_1}}{j_1^{n_1}} \frac{x_2^{j_2}}{j_2^{n_2}} ... \frac{x_k^{j_k}}{j_k^{n_k}}.
\eq
The sum representation gives rise to a quasi-shuffle product and one has for example
\bq
 \mathrm{Li}_{n_1}\left(x_1\right) \mathrm{Li}_{n_2}\left(x_2\right)
 & = &
 \mathrm{Li}_{n_1,n_2}\left(x_1,x_2\right)
 +
 \mathrm{Li}_{n_2,n_1}\left(x_2,x_1\right)
 +
 \mathrm{Li}_{n_1+n_2}\left(x_1 \cdot x_2\right).
\eq
We recall that the Clausen functions are given by
\bq
\mathrm{Cl}_n(\varphi) & = & 
 \left\{ \begin{array}{rl}
   \frac{1}{2i} \left[ \mathrm{Li}_n\left( e^{i \varphi} \right) 
                      -\mathrm{Li}_n\left( e^{-i \varphi} \right)
                \right], 
   & \mbox{$n$ even,} \\
   & \\
   \frac{1}{2} \left[ \mathrm{Li}_n\left( e^{i \varphi} \right) 
                      +\mathrm{Li}_n\left( e^{-i \varphi} \right)
                \right], 
   & \mbox{$n$ odd}, \\
 \end{array} \right.
\eq
while the Glaisher functions are given by
\bq
\mathrm{Gl}_n(\varphi) & = & 
 \left\{ \begin{array}{rl}
   \frac{1}{2} \left[ \mathrm{Li}_n\left( e^{i \varphi} \right) 
                      +\mathrm{Li}_n\left( e^{-i \varphi} \right)
                \right] 
   & \mbox{$n$ even,} \\
   & \\
   \frac{1}{2i} \left[ \mathrm{Li}_n\left( e^{i \varphi} \right) 
                      -\mathrm{Li}_n\left( e^{-i \varphi} \right)
                \right], 
   & \mbox{$n$ odd.} \\
 \end{array} \right.
\eq
The Clausen and Glaisher functions correspond to the real and imaginary part of $\mathrm{Li}_n(e^{i\varphi})$,
and the actual assignment depends on whether $n$ is even or odd.

Let us now consider the elliptic setting.
In \cite{Adams:2014vja} we considered 
the following generalisation depending on three variables $x$, $y$, $q$ and two (integer) indices $n$, $m$:
\bq
 \mathrm{ELi}_{n;m}\left(x;y;q\right) & = & 
 \sum\limits_{j=1}^\infty \sum\limits_{k=1}^\infty \; \frac{x^j}{j^n} \frac{y^k}{k^m} q^{j k}.
\eq
We define the weight of $\mathrm{ELi}_{n;m}(x;y;q)$ to be $w=n+m$.
The definition is symmetric under the exchange of the pair $(x,n)$ with $(y,m)$.
The two summations are coupled through the variable $q$.
In the special case $q=1$ the two summations decouple and we obtain a product of classical polylogarithms:
\bq
 \mathrm{ELi}_{n;m}\left(x;y;1\right) & = & 
 \mathrm{Li}_{n}\left(x\right) \mathrm{Li}_{m}\left(y\right).
\eq
In addition we introduce the following linear combinations
\bq
\label{def_classical_E}
\lefteqn{
 \mathrm{E}_{n;m}\left(x;y;q\right) 
 = } & & \\
 & = &
 \left\{ \begin{array}{ll}
 \frac{1}{i}
 \left[
 \frac{1}{2} \mathrm{Li}_n\left( x \right) - \frac{1}{2} \mathrm{Li}_n\left( x^{-1} \right)
 + \mathrm{ELi}_{n;m}\left(x;y;q\right) - \mathrm{ELi}_{n;m}\left(x^{-1};y^{-1};q\right)
 \right],
 & \mbox{$n+m$ even,} \\
 & \\
 \frac{1}{2} \mathrm{Li}_n\left( x \right) + \frac{1}{2} \mathrm{Li}_n\left( x^{-1} \right)
 + \mathrm{ELi}_{n;m}\left(x;y;q\right) + \mathrm{ELi}_{n;m}\left(x^{-1};y^{-1};q\right),
 & \mbox{$n+m$ odd.} \\
 \end{array}
 \right.
 \nonumber
\eq
The special case $(n,m)=(2,0)$ appeared already in \cite{Adams:2014vja}.
Eq.~(\ref{def_classical_E}) gives the generalisation to arbitrary indices $(n,m)$.
In general, the functions $\mathrm{E}_{n;m}(x;y;q)$ are not symmetric under the exchange of the pair $(x,n)$ with $(y,m)$,
nor do they have for $m\neq 0$ a uniform weight.
The functions $\mathrm{E}_{n;m}(x;y;q)$ can be thought of as elliptic generalisations of the Clausen and Glaisher functions.
In particular we have
\bq
\label{examples_elliptic_generalisations}
 \lim\limits_{q \rightarrow 0} \mathrm{E}_{1;0}\left(e^{i \varphi}; y; q \right)
 & = &
 \mathrm{Cl}_1\left(\varphi\right),
 \nonumber \\
 \lim\limits_{q \rightarrow 0} \mathrm{E}_{2;0}\left(e^{i \varphi}; y; q \right)
 & = &
 \mathrm{Cl}_2\left(\varphi\right),
 \nonumber \\
 \lim\limits_{q \rightarrow 0} \mathrm{E}_{3;1}\left(e^{i \varphi}; y; q \right)
 & = &
 \mathrm{Gl}_3\left(\varphi\right).
\eq
We now turn to the multi-variable case.
In order to keep the notation simple in this case, we introduce a prefactor $c_n$ and a sign $s_n$, both depending on an index $n$ by
\bq
 c_n = \frac{1}{2} \left[ \left(1+i\right) + \left(1-i\right)\left(-1\right)^n\right] = 
 \left\{ \begin{array}{rl}
 1, & \mbox{$n$ even}, \\
 i, & \mbox{$n$ odd}, \\
 \end{array} \right.
 & &
 s_n = (-1)^n =
 \left\{ \begin{array}{rl}
 1, & \mbox{$n$ even}, \\
 -1, & \mbox{$n$ odd}. \\
 \end{array} \right.
\eq
With the help of these definitions we can write the definition of the functions $\mathrm{E}_{n;m}(x;y;q)$
uniformly as
\bq
\lefteqn{
 \mathrm{E}_{n;m}\left(x;y;q\right) 
 = } & & 
 \\
 & = &
 \frac{c_{n+m}}{i}
 \left[
  \left( \frac{1}{2} \mathrm{Li}_n\left( x \right) + \mathrm{ELi}_{n;m}\left(x;y;q\right) \right)
  - s_{n+m}
  \left( \frac{1}{2} \mathrm{Li}_n\left( x^{-1} \right) + \mathrm{ELi}_{n;m}\left(x^{-1};y^{-1};q\right)
  \right)
 \right].
 \nonumber
\eq
It turns out, that in addition to the functions defined above we only need a single depth two elliptic object
in order to express all results of this paper.
This function depends on five variables $x_1$, $x_2$, $y_1$, $y_2$, $q$ and five (integer) indices $n_1$, $n_2$, $m_1$, $m_2$ and $o$.
This new function is defined as follows:
\bq
\lefteqn{
 \mathrm{E}_{n_1,n_2;m_1,m_2;2o}\left(x_1,x_2;y_1,y_2;q\right)
 = 
 } & & \\
 & & 
 \frac{c_{n_1+m_1}}{i} \frac{c_{n_2+m_2}}{i}
 \left\{
 \left[ \mathrm{ELi}_{n_1+o; m_1+o}\left(x_1;y_1;q\right) - s_{n_1+m_1} \mathrm{ELi}_{n_1+o; m_1+o}\left(x_1^{-1};y_1^{-1};q\right) \right]
 \right. \nonumber \\
 & & \left.
 \times
 \frac{1}{2} 
 \left[ \mathrm{Li}_{n_2}\left(x_2\right) - s_{n_2+m_2} \mathrm{Li}_{n_2}\left(x_2^{-1}\right) \right]
 \right. \nonumber \\
 & & \left.
 +
 \sum\limits_{j_1=1}^\infty 
 \sum\limits_{k_1=1}^\infty 
 \sum\limits_{j_2=1}^\infty 
 \sum\limits_{k_2=1}^\infty 
 \frac{\left( x_1^{j_1} y_1^{k_1} - s_{n_1+m_1} x_1^{-j_1} y_1^{-k_1} \right)}{j_1^{n_1} k_1^{m_1}}
 \frac{\left( x_2^{j_2} y_2^{k_2} - s_{n_2+m_2} x_2^{-j_2} y_2^{-k_2} \right)}{j_2^{n_2} k_2^{m_2}}
 \frac{q^{j_1 k_1 + j_2 k_2}}{\left(j_1 k_1 + j_2 k_2\right)^o}
 \right\}.
 \nonumber
\eq
Please note that this definition is asymmetric in the quadruplets $(n_1,m_1,x_1,y_1)$ and $(n_2,m_2,x_2,y_2)$. 
Let us briefly discuss the weights of the individual pieces. The first term, consisting of products of the form
\bq
 \mathrm{ELi}_{n_1+o; m_1+o}\left(x_1;y_1;q\right) \mathrm{Li}_{n_2}\left(x_2\right)
\eq
is of weight $w_1=n_1+n_2+m_1+2o$, while the quadruple sum of the form
\bq
 \sum\limits_{j_1=1}^\infty 
 \sum\limits_{k_1=1}^\infty 
 \sum\limits_{j_2=1}^\infty 
 \sum\limits_{k_2=1}^\infty 
 \frac{ x_1^{j_1} y_1^{k_1} }{j_1^{n_1} k_1^{m_1}}
 \frac{ x_2^{j_2} y_2^{k_2} }{j_2^{n_2} k_2^{m_2}}
 \frac{q^{j_1 k_1 + j_2 k_2}}{\left(j_1 k_1 + j_2 k_2\right)^o}
\eq
is of weight $w_2=n_1+n_2+m_1+m_2+2o$.
For $m_2=0$ the two weights coincide.

For $o > 0$ we can express this function as an $o$-fold iterated integral over $q$ (the variables $x_1$, $x_2$, $y_1$ and $y_2$
are treated as constants in the integration):
\bq
\label{E_depth_two_integral_repr}
\lefteqn{
 \mathrm{E}_{n_1,n_2;m_1,m_2;2o}\left(x_1,x_2;y_1,y_2;q\right)
 = 
 } & & \\
 & &
 \int\limits_0^q \frac{dq_1}{q_1} \int\limits_0^{q_1} \frac{dq_2}{q_2} ... \int\limits_0^{q_{o-1}} \frac{dq_o}{q_o}
 \left[ \mathrm{E}_{n_1; m_1}\left(x_1;y_1;q_o\right) - \mathrm{E}_{n_1; m_1}\left(x_1;y_1;0\right) \right]
 \mathrm{E}_{n_2;m_2}\left(x_2;y_2;q_o\right).
 \nonumber
\eq
In eq.~(\ref{E_depth_two_integral_repr}) the asymmetry with respect to the quadruplets $(n_1,m_1,x_1,y_1)$ and $(n_2,m_2,x_2,y_2)$
manifests itself by the fact, that the constant part with respect to the Taylor expansion in $q$ is subtracted out
from $\mathrm{E}_{n_1; m_1}(x_1;y_1;q)$, but not from $\mathrm{E}_{n_2;m_2}(x_2;y_2;q)$.
This subtraction ensures that the integrand has a Taylor expansion starting at $q^1$.
Therefore the integral is well defined at the lower integration boundary $q=0$ and no regularisation is needed.
 
Let us write an explicit example. At depth two there is only one combination of indices, which 
will appear in the results of this paper.
The indices are given by $(n_1,n_2)=(0,1)$, $(m_1,m_2)=(-2,0)$ and $2o=4$.
We have
\bq
\lefteqn{
 \mathrm{E}_{0,1;-2,0;4}\left(x_1,x_2;y_1,y_2;-q\right)
 = 
 } & & \\
 & & 
 \frac{1}{i}
 \left\{
 \left[ \mathrm{ELi}_{2; 0}\left(x_1;y_1;-q\right) - \mathrm{ELi}_{2; 0}\left(x_1^{-1};y_1^{-1};-q\right) \right]
 \times
 \frac{1}{2} 
 \left[ \mathrm{Li}_{1}\left(x_2\right) + \mathrm{Li}_{1}\left(x_2^{-1}\right) \right]
 \right. \nonumber \\
 & & \left.
 +
 \sum\limits_{j_1=1}^\infty 
 \sum\limits_{k_1=1}^\infty 
 \sum\limits_{j_2=1}^\infty 
 \sum\limits_{k_2=1}^\infty 
 \frac{k_1^2}{j_2 \left(j_1 k_1 + j_2 k_2\right)^2}
 \left( x_1^{j_1} y_1^{k_1} - x_1^{-j_1} y_1^{-k_1} \right)
 \left( x_2^{j_2} y_2^{k_2} + x_2^{-j_2} y_2^{-k_2} \right)
 \left(-q\right)^{j_1 k_1 + j_2 k_2}
 \right\}.
 \nonumber
\eq
The corresponding integral representation reads
\bq
\lefteqn{
 \mathrm{E}_{0,1;-2,0;4}\left(x_1,x_2;y_1,y_2;-q\right)
 = 
 } & & \\
 & & 
 \int\limits_0^{q} \frac{dq_1}{q_1} \int\limits_0^{q_1} \frac{dq_2}{q_2} 
 \left[ \mathrm{E}_{0; -2}\left(x_1;y_1;-q_2\right) - \mathrm{E}_{0; -2}\left(x_1;y_1;0\right) \right]
 \mathrm{E}_{1; 0}\left(x_2;y_2;-q_2\right).
 \nonumber
\eq

\section{The result for $S_{111}^{(1)}(2,t)$}
\label{sec:result_S1}

In this section we present the ${\mathcal O}(\eps^1)$-term $S_{111}^{(1)}(2,t)$ of the sunrise integral around two space-time
dimensions in terms of elliptic generalisations of the Clausen and Glaisher functions discussed in the previous section.

Let us start by recalling that the ${\mathcal O}(\eps^0)$-term $S_{111}^{(0)}(2,t)$ of the $\eps$-expansion of
$S_{111}(2-2\eps,t)$ is given by
\bq
 S_{111}^{(0)}\left(2,t\right)
 & = &
 \frac{\psi_1}{\pi} E^{(0)},
\eq
with 
\bq
\label{res_E_0}
 E^{(0)}
 & = &
 \mathrm{E}_{2;0}\left(w_1;-1;-q\right)
 +
 \mathrm{E}_{2;0}\left(w_2;-1;-q\right)
 +
 \mathrm{E}_{2;0}\left(w_3;-1;-q\right).
\eq
This motivates the following ansatz for $S_{111}^{(1)}(2,t)$:
\bq
 S_{111}^{(1)}\left(2,t\right)
 & = &
 \frac{\psi_1}{\pi} E^{(1)}.
\eq
We find that $E^{(1)}$ is given by
\bq
\label{res_E_1}
\lefteqn{
 E^{(1)}
 = 
} & &
 \nonumber \\
 & &
 \left\{ 
         - \frac{2}{3} \sum \limits_{j=1}^3  \ln\left( \frac{m_j^2}{\mu^2} \right) 
         - 6 \mathrm{E}_{1,0}\left(-1;1;-q\right)
         + \sum\limits_{j=1}^3 \left[ 
                                      \mathrm{E}_{1;0}\left(w_j;1;-q\right) 
                                      - \frac{1}{3} \mathrm{E}_{1;0}\left(w_j;-1;-q\right) 
                               \right]
 \right\} E^{(0)}
 \nonumber \\
 & &
 - 2 
 \sum\limits_{j=1}^3
 \frac{1}{2i} 
  \left\{ 
           \mathrm{Li}_{2,1}\left(w_j,1\right) - \mathrm{Li}_{2,1}\left(w_j^{-1},1\right)
           + \mathrm{Li}_3\left(w_j\right) - \mathrm{Li}_3\left(w_j^{-1}\right) 
 \right. \nonumber \\
 & & \left.
           + 3 \ln\left(2\right) \left[ 
                                        \mathrm{Li}_2\left( w_j \right) 
                                        -\mathrm{Li}_2\left( w_j^{-1} \right)
                                 \right] 
  \right\}
 \nonumber \\
 & &
 + 
 \sum\limits_{j=1}^3
   \left[ 
          4 \mathrm{E}_{0,1;-2,0;4}\left( w_j,w_j; -1,-1; -q \right)
          -
          6 \mathrm{E}_{0,1;-2,0;4}\left( w_j,w_j; 1,-1; -q \right)
 \right. \nonumber \\
 & & \left.
          +
          6 \mathrm{E}_{0,1;-2,0;4}\left( w_j,-1; -1,1; -q \right)
   \right]
 \nonumber \\
 & &
 + 
 \sum\limits_{j_1=1}^3
 \sum\limits_{j_2=1}^3
  \left[
         2 \mathrm{E}_{0,1;-2,0;4}\left( w_{j_1},w_{j_2}; 1,-1; -q \right)
         - \mathrm{E}_{0,1;-2,0;4}\left( w_{j_1},w_{j_2}; -1,-1; -q \right) 
 \right. \nonumber \\
 & & \left.
         - \mathrm{E}_{0,1;-2,0;4}\left( w_{j_1},w_{j_2}; -1,1; -q \right) 
  \right]
 \nonumber \\
 & &
 +
 \sum\limits_{j=1}^3 \mathrm{E}_{3;1}\left(w_j;-1;-q\right).
\eq
Eq.~(\ref{res_E_1}) is one of the main results of this paper.
A few remarks are in order:
Eq.~(\ref{res_E_1}) gives the result for $E^{(1)}$ entirely in the variables $q$, $w_1$, $w_2$, $w_3$
and $m_1^2 m_2^2 m_3^2 / \mu^6$.
Although not as elegant as the result for $E^{(0)}$ in eq.~(\ref{res_E_0}), it is still far from trivial
that $E^{(1)}$ can be expressed in a few lines.
As arguments of the (elliptic) multiple polylogarithms only the variables $w_j^{\pm 1}$ and the values $\pm 1$ occur.

The first line gives all terms proportional to the ${\mathcal O}(\eps^0)$ result $E^{(0)}$.
These terms are all of weight one, multiplying $E^{(0)}$, which is of weight two, yielding the total weight three.
The second and third line contain terms, which depend on the variables $w_j$, but not explicitly on $q$.
These are ordinary multiple polylogarithms.
Terms like these are expected, 
since $S_{111}^{(1)}(2,t)$ depends not only on the graph polynomial ${\mathcal F}$, but also
on the graph polynomial ${\mathcal U}$.
The ordinary multiple polylogarithms are all of weight three.
The next four lines contain functions of depth two.
Surprisingly, we only encounter the function $\mathrm{E}_{0,1;-2,0;4}$, which is of weight $1-2+4=3$.
The last line gives probably the most interesting part: 
It contains the elliptic polylogarithm $\mathrm{E}_{3;1}$,
an elliptic generalisation of the Glaisher function $\mathrm{Gl}_3$ according to eq.~(\ref{examples_elliptic_generalisations}).
The elliptic polylogarithm $\mathrm{E}_{3;1}$ is not homogeneous in the weight,
having parts of weight $3$ and parts of weight $4$.
The occurrence of $\mathrm{E}_{3;1}$ is discussed in more detail in section \ref{sec:weights}.

Furthermore, the result shows an explicit $\ln(2)$. These $\ln(2)$-terms are spurious and cancel with
$\ln(2)$-terms from $\mathrm{E}_{1;0}(-1;1;-q)$ and
$\mathrm{E}_{0,1;-2,0;4}( w_j,-1; -1,1; -q )$.

\section{The result for the sunrise integral around four space-time dimensions}
\label{sec:shift}

In this section we discuss dimensional-shift relations, which relate the integrals $S_{111}(4-2\eps,t)$ and 
$S_{111}(2-2\eps,t)$.
This allows us to express the finite part $S_{111}^{(0)}(4,t)$ of the sunrise integral around four space-time dimensions
in terms of the already calculated integrals $S_{111}^{(0)}(2,t)$ and $S_{111}^{(1)}(2,t)$.
The divergent parts $S_{111}^{(-2)}(4,t)$ and $S_{111}^{(-1)}(4,t)$ are rather simple and listed for completeness.

There are several relations, which relate integrals in $D$ space-time dimensions 
to integrals in $(D+2)$ space-time dimensions.
The Tarasov relations read \cite{Tarasov:1996br,Tarasov:1997kx}
\bq
 S_{\nu_1 \nu_2 \nu_3}\left( D, t\right)
 & = &
  \nu_1 \nu_2 S_{(\nu_1+1) (\nu_2+1) \nu_3}\left( D+2, t\right)
  +
  \nu_2 \nu_3 S_{\nu_1 (\nu_2+1) (\nu_3+1)}\left( D+2, t\right)
 \nonumber \\
 & &
  +
  \nu_1 \nu_3 S_{(\nu_1+1) \nu_2 (\nu_3+1)}\left( D+2, t\right).
\eq
From eq.~(\ref{def_Feynman_integral}) we find
\bq
 \mu^2 \frac{d}{dt}
 S_{\nu_1 \nu_2 \nu_3}\left( D, t\right)
 & = &
 \nu_1 \nu_2 \nu_3
 S_{(\nu_1+1) (\nu_2+1) (\nu_3+1)}\left( D+2, t\right).
\eq
For a fixed space-time dimension we can always reduce any integral to a linear combination of sunrise master integrals
and simpler integrals.
Inverting the relations, which relate integrals in $D$ dimensions to integrals in $(D+2)$ dimensions, we can express
$S_{111}(4-2\eps,t)$ in terms of four sunrise master integrals for $D=2-2\eps$ and simpler tadpole integrals.
A possible basis of sunrise master integrals is given by 
\bq
\label{derivative_basis}
 S_{111}\left(D,t\right),
 \;\;\;\;\;\;
 \mu^2 \frac{d}{dt} S_{111}\left(D,t\right),
 \;\;\;\;\;\;
 \mu^4 \frac{d^2}{dt^2} S_{111}\left(D,t\right),
 \;\;\;\;\;\;
 \mu^6 \frac{d^3}{dt^3} S_{111}\left(D,t\right).
\eq
In this basis one finds for $S_{111}^{(0)}(4,t)$
\bq
\label{S4_dimensional_shift}
 S_{111}^{(0)}(4,t)
 & = &
 \frac{1}{\mu^4}
 L_3^{(-1)}\left(2\right) \;\; S_{111}^{(1)}(2,t)
 +  \frac{1}{\mu^4} L_3^{(0)}\left(2\right) \;\; S_{111}^{(0)}(2,t)
 +
 R^{(0)},
\eq
where the remainder function $R^{(0)}$ contains the contributions from the tadpoles.
The differential operator $L_3^{(-1)}(2)$ factorises again:
\bq
 L_3^{(-1)}\left(2\right)
 & = &
 L^{(-1)}_{1}\left(2\right) 
 \;\;
 L_2^{(0)}\left(2\right),
\eq
with $L^{(-1)}_{1}(2)$ given by
\bq
\label{def_L_1_m1}
 L^{(-1)}_{1}\left(2\right)
 & = &
 \frac{-33t^3+13 M_{100}t^2 + \left(22 M_{110}-15 M_{200} \right)t - 3 M_{100} \Delta}
      {24 t \left(3t^2-2 M_{100}t+\Delta\right) \left(5t^2 + 2M_{100} t + 7 \Delta \right)}
 \frac{d}{dt}
 \nonumber \\
 & &
 +
 \frac{69 t^2 - 22 M_{100} t + 9 M_{200} -10 M_{110}}{24 t \left(3t^2-2 M_{100}t+\Delta\right) \left(5t^2 + 2M_{100} t + 7 \Delta \right)}.
\eq
The explicit expressions for the differential operator $L_3^{(0)}(2)$ and the remainder function $R^{(0)}$ are rather long and not
provided here.
We have used eq.~(\ref{S4_dimensional_shift}) internally for the determination of the integration constants $C_1$ and $C_2$.

In order to present our final result on $S_{111}^{(0)}(4,t)$ it is better to use an alternative basis of sunrise master integrals
given by
\bq
 S_{111}\left(D,t\right),
 \;\;\;\;\;\;
 \mu^2 \frac{\partial}{\partial m_1^2} S_{111}\left(D,t\right),
 \;\;\;\;\;\;
 \mu^2 \frac{\partial}{\partial m_2^2} S_{111}\left(D,t\right),
 \;\;\;\;\;\;
 \mu^2 \frac{\partial}{\partial m_3^2} S_{111}\left(D,t\right).
\eq
This is the basis of sunrise master integrals used in \cite{Caffo:1998du}.
We recall that
\bq
 \mu^2 \frac{\partial}{\partial t}
 S_{111}\left( D, t\right)
 & = &
 S_{222}\left( D+2, t\right),
 \nonumber \\
 \mu^2 \frac{\partial}{\partial m_1^2}
 S_{111}\left( D, t\right)
 & = &
 - S_{211}\left( D, t\right),
 \nonumber \\
 \mu^2 \frac{\partial}{\partial m_2^2}
 S_{111}\left( D, t\right)
 & = &
 - S_{121}\left( D, t\right),
 \nonumber \\
 \mu^2 \frac{\partial}{\partial m_3^2}
 S_{111}\left( D, t\right)
 & = &
 - S_{112}\left( D, t\right).
\eq
Let us define for $k \in \{-1,0\}$ two partial differential operators
\bq
 \tilde{L}_3^{(k)}(2) & = & 
 C^{(k)}_0
 +
 \sum\limits_{i=1}^3 C^{(k)}_i m_i^2 \frac{\partial}{\partial m_i^2}.
\eq
For $k=-1$ the coefficients are given by
\bq
 C^{(-1)}_i & = &
 \frac{\left(t-m_i^2\right)}{6 \mu^4 t} \left[ \left( 2 m_i^2 - m_j^2 - m_k^2 \right) t 
                           - 3 m_j^4 - 3 m_k^4 - m_i^2 m_j^2 - m_i^2 m_k^2 + 8 m_j^2 m_k^2 \right],
 \nonumber \\
 C^{(-1)}_0 & = &
 \frac{1}{4} \left( C^{(-1)}_1 + C^{(-1)}_2 + C^{(-1)}_3 \right),
\eq
while for $k=0$ we have
\bq
 C^{(0)}_0 & = &
 \frac{1}{12 \mu^4 t}
 \left[
        - M_{100} t^2 + \left( -26M_{200}+36M_{110} \right)t + M_{300} + 8 M_{210} - 78 M_{111}
 \right]
 \nonumber \\
 C^{(0)}_i & = &
 \frac{\left(t-m_i^2\right)}{12 \mu^4 t} \left[ -t ^2 + \left( 12 m_i^2 - 3 m_j^2 - 3 m_k^2 \right) t 
                           - m_i^4 - 16 m_j^4 - 16 m_k^4 - 3 m_i^2 m_j^2 - 3 m_i^2 m_k^2 
 \right. \nonumber \\
 & & \left.
 + 66 m_j^2 m_k^2 \right].
\eq
Please note that all coefficients of $\tilde{L}_3^{(-1)}(2)$ vanish in the equal mass case.
For $S_{111}^{(0)}\left(4,t\right)$ we find 
\bq
\label{res_S4}
\lefteqn{
 S_{111}^{(0)}\left(4,t\right)
 = 
 \tilde{L}_3^{(-1)}(2) S_{111}^{(1)}\left(2,t\right)
 +
 \tilde{L}_3^{(0)}(2) S_{111}^{(0)}\left(2,t\right)
 + \frac{13}{8} \frac{t}{\mu^2} - 3 \frac{M_{100}}{\mu^2} - \frac{1}{2} \zeta_2 \frac{M_{100}}{\mu^2} 
} & &
 \\
 & &
 - \sum\limits_{(i,j,k) \in {\mathbb Z}_3} 
     \left[ 
           \frac{\left(4m_i^2 + m_j^2 + m_k^2\right) t + m_i^2\left(m_j^2+m_k^2\right) - 2 m_j^2 m_k^2}{12 \mu^2 t} \ln^2\left(\frac{m_i^2}{\mu^2}\right) 
 \right. \nonumber \\
 & &
 \left.
 - \frac{\left(m_i^2 - 2m_j^2 - 2m_k^2\right) t + m_i^2\left(m_j^2+m_k^2\right) - 2 m_j^2 m_k^2}{6 \mu^2 t} \ln\left(\frac{m_j^2}{\mu^2}\right) \ln\left(\frac{m_k^2}{\mu^2}\right) 
 \right. \nonumber \\
 & &
 \left. 
 + \frac{2 t^2 - \left(24m_i^2 + 6m_j^2 + 6m_k^2\right) t + 2m_i^4 - m_j^4 - m_k^4 - 6m_i^2\left(m_j^2+m_k^2\right) + 12 m_j^2 m_k^2}{12 \mu^2 t} \ln\left(\frac{m_i^2}{\mu^2}\right)
 \right].
 \nonumber
\eq
The notation $(i,j,k) \in {\mathbb Z}_3$ stands
for a sum over the three cyclic permutations of $(1,2,3)$.
Eq.~(\ref{res_S4}) expresses the finite part $S_{111}^{(0)}(4,t)$ of the sunrise integral in $4-2\eps$ dimensions
in terms of $S_{111}^{(1)}(2,t)$ and $S_{111}^{(0)}(2,t)$, derivatives thereof and simpler terms.
The results for $S_{111}^{(1)}(2,t)$ and $S_{111}^{(0)}(2,t)$ have been given in section~\ref{sec:result_S1}.
Please note that $\tilde{L}_3^{(-1)}(2)$ vanishes in the equal mass case.
Therefore it follows that in the equal mass case the ${\mathcal O}(\eps)$-part $S_{111}^{(1)}(2,t)$ does not affect
$S_{111}^{(0)}(4,t)$.
However, in the unequal mass case the ${\mathcal O}(\eps)$-part $S_{111}^{(1)}(2,t)$ is required
for $S_{111}^{(0)}(4,t)$.
Eq.~(\ref{res_S4}) together with eq.~(\ref{res_E_1}) is the main result of this paper.

For completeness we list also the divergent terms of $S_{111}(4-2\eps,t)$. These read
\bq
 S_{111}^{(-2)}(4,t)
 & = &
 - \frac{M_{100}}{2 \mu^2},
 \nonumber \\
 S_{111}^{(-1)}(4,t)
 & = &
 \frac{t}{4 \mu^2} 
          - \frac{3 M_{100}}{2 \mu^2} 
          + \sum\limits_{i=1}^3 \frac{m_i^2}{\mu^2} \ln\left(\frac{m_i^2}{\mu^2}\right).
\eq

\section{The equal mass case}
\label{sec:equal_mass_case}

This section is devoted to the equal mass case $m_1=m_2=m_3=m$.
In the equal mass case the result for the sunrise integral around four space-time dimensions is significantly
simpler.
We have already seen that the contribution from $S_{111}^{(1)}(2,t)$ in 
eq.~(\ref{res_S4}) drops out in the equal mass case.
Furthermore we have now a second-order differential equation for all $D$ \cite{Laporta:2004rb}:
\bq
\lefteqn{
 \left\{
 2 t \left(t-9m^2\right) \left(t-m^2\right) \frac{d^2}{dt^2}
 + \left[ 3 \left(4-D\right) t^2 + 10 \left(D-6\right) t m^2 + 9 D m^4 \right] \frac{d}{dt}
 \right.
 } & & \nonumber \\
 & &
 \left.
 + \left(D-3\right) \left[ \left(D-4\right) t + \left(D+4\right) m^2 \right]
 \right\}
 S_{111}\left(D,t\right)
 = 
 - 3 \left(D-2\right)^2 \mu^2 \left[ T_1\left(D\right) \right]^2,
\eq
with $T_1(D)=T_1(D,m^2,\mu^2)$.
A convenient basis of sunrise master integrals is therefore
\bq
 S_{111}\left(D,t\right),
 \;\;\;\;\;\;
 \mu^2 \frac{d}{dt} S_{111}\left(D,t\right).
\eq
In terms of this basis the dimensional-shift relation simplifies and we have
\bq
 S_{111}\left(4-2\eps,t\right)
 & = &
 \frac{1}{6 \left(1-2\eps\right)\left(1-3\eps\right)\left(2-3\eps\right)}
 \left\{
         \frac{\left(t+3m^2\right)\left(t-m^2\right)\left(t-9m^2\right)}{\mu^4} \frac{d}{dt} S_{111}\left(2-2\eps,t\right)
 \right. \nonumber \\
 & & \left.
         + \left[ \frac{\left(t-m^2\right)\left(t-9m^2\right)}{\mu^4} 
                + \eps \frac{\left( t^2 + 22 m^2 t - 87 m^4 \right)}{\mu^4} \right] S_{111}\left(2-2\eps,t\right)
 \right. \nonumber \\
 & & \left.
         + 3 \left(1-\eps\right)^2 \left[ - \frac{6 \mu^2}{m^2} + \eps \frac{\mu^2 \left( t + 21 m^2 \right)}{m^4} \right] \left[ T_1\left(4-2\eps\right) \right]^2
 \right\}.
\eq
For the coefficients of the $\eps$-expansion we have now
\bq
 S_{111}^{(-2)}(4,t)
 & = &
 - \frac{3 m^2}{2 \mu^2},
 \\
 S_{111}^{(-1)}(4,t)
 & = &
 \frac{t}{4 \mu^2} 
          - \frac{9 m^2}{2 \mu^2} 
          + \frac{3 m^2}{\mu^2} \ln\left(\frac{m^2}{\mu^2}\right),
 \nonumber \\
 S_{111}^{(0)}(4,t)
 & = &
 \frac{13 t}{8 \mu^2}
 - \frac{9 m^2}{\mu^2}
 - \frac{3 m^2}{2 \mu^2} \zeta_2
 - \left( \frac{t}{2\mu^2} - \frac{9 m^2}{\mu^2} \right) \ln\left(\frac{m^2}{\mu^2}\right)
 - \frac{3 m^2}{\mu^2} \left(\ln\left(\frac{m^2}{\mu^2}\right) \right)^2
 \nonumber \\
 & &
 + \frac{\left(t-m^2\right)\left(t-9m^2\right)}{12 \mu^4} S_{111}^{(0)}\left(2,t\right)
 + \frac{\left(t+3m^2\right)\left(t-m^2\right)\left(t-9m^2\right)}{12 \mu^4} \frac{d}{dt} S_{111}^{(0)}\left(2,t\right),
 \nonumber
\eq
again showing the absence of $S_{111}^{(1)}(2,t)$ in the expression for $S_{111}^{(0)}(4,t)$.

\section{Weights}
\label{sec:weights}

In this section we discuss in more detail the transcendental weights associated with individual terms and in particular the
occurrence of the functions $\mathrm{E}_{3;1}$.

The building blocks of all our expressions are the multiple polylogarithms of the form
\bq
\label{example_polylogs}
 \mathrm{Li}_{n_1,n_2,...,n_k}\left(x_1,x_2,...,x_k\right)
 & = &
 \sum\limits_{j_1=1}^\infty \sum\limits_{j_2=1}^{j_1-1} ... \sum\limits_{j_k=1}^{j_{k-1}-1}
 \frac{x_1^{j_1}}{j_1^{n_1}} \frac{x_2^{j_2}}{j_2^{n_2}} ... \frac{x_k^{j_k}}{j_k^{n_k}},
\eq
the generalisation of the classical polylogarithms to the elliptic case
\bq
\label{example_elliptic_polylogs}
 \mathrm{ELi}_{n_1;m_1}\left(x_1;y_1;q\right) & = & 
 \sum\limits_{j_1=1}^\infty \sum\limits_{k_1=1}^\infty \; \frac{x_1^{j_1}}{j_1^{n_1}} \frac{y_1^{k_1}}{k_1^{m_1}} q^{j_1 k_1},
\eq
and quadruple sums of the form
\bq
\label{example_quadruple_sum}
 \sum\limits_{j_1=1}^\infty 
 \sum\limits_{k_1=1}^\infty 
 \sum\limits_{j_2=1}^\infty 
 \sum\limits_{k_2=1}^\infty 
 \frac{ x_1^{j_1} y_1^{k_1} }{j_1^{n_1} k_1^{m_1}}
 \frac{ x_2^{j_2} y_2^{k_2} }{j_2^{n_2} k_2^{m_2}}
 \frac{q^{j_1 k_1 + j_2 k_2}}{\left(j_1 k_1 + j_2 k_2\right)^o}.
\eq
For $x_1=x_2=...x_k=1$, $y_1=y_2=1$ and $q=1$ the summand is in all cases a homogeneous function of the summation variables.
We defined the transcendental weight as the negative of the degree of homogeneity with respect to the summation variables in this
case.
Thus eq.~(\ref{example_polylogs}) is of weight $w_1=n_1+n_2+...+n_k$,
eq.~(\ref{example_elliptic_polylogs}) is of weight $w_2=n_1+m_1$, and
eq.~(\ref{example_quadruple_sum}) is of weight $w_3=n_1+n_2+m_1+m_2+2 o$.

With this weight counting, the function
\bq
 \mathrm{ELi}_{3;1}\left(x_1;y_1;q\right)
\eq
is of weight $4$ and we would like to discuss the occurrence of this function in the result for $S_{111}^{(1)}(2,t)$.
It turns out, that the occurrence of this function can be related to the fact that
\bq
 \ln q
\eq
should be counted as weight $2$.
The latter fact can be seen as follows: We have for example
\bq
 \int\limits_0^x \mathrm{ELi}_{n;m}\left(x';y;q\right) d \ln x'
 & = & 
 \mathrm{ELi}_{n+1;m}\left(x;y;q\right),
\eq
but
\bq
 \int\limits_0^q \mathrm{ELi}_{n;m}\left(x;y;q'\right) d \ln q'
 & = & 
 \mathrm{ELi}_{n+1;m+1}\left(x;y;q\right).
\eq
Thus an integration with respect to $d\ln q = dq / q$ increases the weight by $2$.
The final result for $S_{111}^{(1)}(2,t)$ does not contain any $\ln q$-terms, however a convenient method
for calculating $S_{111}^{(1)}(2,t)$ splits the integral into two parts, each part containing $\ln q$-terms.
The $\ln q$-terms cancel in the sum, but leave the functions $\mathrm{ELi}_{3;1}$ as remainders.

In order to understand how this happens, it is sufficient to consider the equal mass case $m_1=m_2=m_3=m$ for
$S_{111}^{(1)}(2,t)$.
In the equal mass case the differential equations for $S_{111}^{(1)}(2,t)$ (and $S_{111}^{(0)}(2,t)$) simplify and we find 
with
\bq
\lefteqn{
 L^{(0)}_{2,\mathrm{equal}}
 = 
 p_{2,\mathrm{equal}} \frac{d^2}{dt^2}
 +
 p_{1,\mathrm{equal}} \frac{d}{dt}
 +
 p_{0,\mathrm{equal}},
 } & &
 \\
 & &
 p_{2,\mathrm{equal}} \; = \; t \left(t-m^2\right) \left(t-9m^2\right),
 \;\;\;\;\;\;
 p_{1,\mathrm{equal}} \; = \; 3 t^2 - 20 t m^2 + 9 m^4,
 \;\;\;\;\;\;
 p_{0,\mathrm{equal}} \; = \; t- 3 m^2,
 \nonumber
\eq
the following differential equations
\bq
 L^{(0)}_{2,\mathrm{equal}} S_{111}^{(0)}\left(2,t\right)
 & = &
 - 6 \mu^2,
 \\
 L^{(0)}_{2,\mathrm{equal}} S_{111}^{(1)}\left(2,t\right)
 & = &
 12 \mu^2 \ln\left(\frac{m^2}{\mu^2}\right)
 + 
 \left[
       \left(-3t^2 + 10 t m^2 + 9 m^4 \right) \frac{d}{dt}  -3t + 5m^2 
 \right] S_{111}^{(0)}\left(2,t\right).
 \nonumber
\eq
In order to find $S_{111}^{(1)}(2,t)$, let us make the ansatz
that $S_{111}^{(1)}(2,t)$ consists of a part proportional to $S_{111}^{(0)}(2,t)$ and a remainder $\tilde{S}_{111}^{(1)}(2,t)$:
\bq
 S_{111}^{(1)}\left(2,t\right)
 & = & 
 \tilde{S}_{111}^{(1)}\left(2,t\right)
 +
 F_1(t) S_{111}^{(0)}\left(2,t\right).
\eq
The differential equation for $\tilde{S}_{111}^{(1)}(2,t)$ is then
\bq
\label{diff_eq_S1_equal_mass_case}
 L^{(0)}_{2,\mathrm{equal}} \tilde{S}_{111}^{(1)}\left(2,t\right)
 & = &
 12 \mu^2 \ln\left(\frac{m^2}{\mu^2}\right)
 + 
 6 \mu^2 F_1\left(t\right)
 -
 \left[
       \left(2 p_{2,\mathrm{equal}} \frac{d F_1(t)}{dt} +3t^2 - 10 t m^2 - 9 m^4 \right) \frac{d}{dt}  
 \right.
 \nonumber \\
 & &
 \left.
       + p_{2,\mathrm{equal}} \frac{d^2F_1(t)}{dt^2} 
       + p_{1,\mathrm{equal}} \frac{dF_1(t)}{dt} 
       + 3t - 5m^2  
 \right] S_{111}^{(0)}\left(2,t\right).
\eq
We are free to choose $F_1(t)$ 
(different choices for $F_1(t)$ will lead to different expressions for $\tilde{S}_{111}^{(1)}(2,t)$)
and a convenient choice for $F_1(t)$ will be the one, which eliminates
$d S_{111}^{(0)}(2,t)/dt$ on the right-hand side of eq.~(\ref{diff_eq_S1_equal_mass_case}).
This amounts to the choice
\bq
 \frac{d}{dt} F_1(t) 
 & = &
 \frac{-3t^2 + 10 t m^2 + 9 m^4}{2 t \left(t-m^2\right)\left(t-9m^2\right)}
 \;\; = \;\;
 \frac{1}{2t} - \frac{1}{t-m^2} - \frac{1}{t-9m^2},
\eq
and hence
\bq
 F_1(t)
 & = &
 - \ln\left( \frac{m \left(t-m^2\right)\left(t-9m^2\right)}{3 \mu^4 \sqrt{t}} \right).
\eq
It turns out that this choice does not only eliminate the terms proportional to the derivative of
$S_{111}^{(0)}(2,t)$, but also the terms proportional to $S_{111}^{(0)}(2,t)$ itself
and the differential equation for $\tilde{S}_{111}^{(1)}(2,t)$ reads
\bq
\label{simplified_dgl}
 L^{(0)}_{2,\mathrm{equal}} \tilde{S}_{111}^{(1)}\left(2,t\right)
 & = &
 \mu^2 \tilde{I}_{2,\mathrm{equal}}\left(t\right),
 \;\;\;\;\;\;\;\;\;
 \tilde{I}_{2,\mathrm{equal}}\left(t\right)
 \;\; = \;\;
 6 \ln\left(\frac{3 m^3 \sqrt{t}}{\left(t-m^2\right)\left(t-9m^2\right)} \right).
\eq
The inhomogeneous term of this differential equation is significantly simpler 
than the generic inhomogeneous term of eq.~(\ref{diff_eq_S1_equal_mass_case}).
But please note that $F_1(t)$ and $\tilde{I}_{2,\mathrm{equal}}(t)$ contain both terms
proportional to $\ln(t)$ (which can be related to $\ln(-q)$-terms).

In order to determine $\tilde{S}_{111}^{(1)}(2,t)$ from eq.~(\ref{simplified_dgl})
one needs first the homogeneous solutions. These are spanned by $\psi_1$ and $\psi_2$, defined
in eq.~(\ref{def_periods}).
For the case at hand it will be convenient to use instead of the basis $\{\psi_1,\psi_2\}$
the basis given by
\bq
 \psi_1,
 & &
 \psi_1 \ln\left(-q\right).
\eq
We then write the full solution as
\bq
\tilde{S}_{111}^{(1)}\left(2,t\right)
 & = & 
 c_1 \psi_1
 + c_2 \psi_1 \ln\left(-q\right) 
 + \tilde{S}_{111,\mathrm{special}}^{(1)}\left(2,t\right)
\eq
with $\tilde{S}_{111,\mathrm{special}}^{(1)}(2,0)=0$.
From the boundary values one finds now
\bq
 c_1 
 & = &
 \frac{3}{2 \pi i} 
 \left\{
  \mathrm{Li}_3\left(r_3\right) - \mathrm{Li}_3\left(r_3^{-1}\right) 
  - 2 \left[ \mathrm{Li}_{21}\left(r_3,1\right) + \mathrm{Li}_3\left(r_3\right) - \mathrm{Li}_{21}\left(r_3^{-1},1\right) - \mathrm{Li}_3\left(r_3^{-1}\right) \right] 
 \right. \nonumber \\
 & & \left.
  - \ln\left(3\right) \left( \mathrm{Li}_2\left(r_3\right) - \mathrm{Li}_2\left(r_3^{-1}\right) \right)
 \right\},
 \nonumber \\
 c_2
 & = &
 - \frac{1}{2} \frac{3}{2 \pi i} \left[ \mathrm{Li}_2\left(r_3\right) - \mathrm{Li}_2\left(r_3^{-1}\right) \right].
\eq
In the following we will denote by
\bq
 r_p & = & \exp\left(\frac{2\pi i}{p}\right)
\eq
the $p$-th root of unity.
For $\tilde{S}_{111,\mathrm{special}}^{(1)}(2,t)$ one finds 
along the lines leading to eq.~(\ref{solution_S1_quadrature})
\bq
 \tilde{S}_{111,\mathrm{special}}^{(1)}\left(2,t\right)
 = 
 -
 \frac{\psi_1}{\pi}
 \frac{1}{2i}
 \sum\limits_{j=1}^\infty
 \sum\limits_{k=1}^\infty
 \int\limits_0^q \frac{dq_1}{q_1}
 \int\limits_0^{q_1} \frac{dq_2}{q_2}
 \left(r_3^j - r_3^{-j} \right) 
 k^2 \left(-1\right)^k
 \left(-q_2\right)^{jk}
 \tilde{I}_{2,\mathrm{equal}}(q_2).
\eq
In order to perform the integration we need the $q$-expansion of $\tilde{I}_{2,\mathrm{equal}}$.
The $q$-expansion of $\tilde{I}_{2,\mathrm{equal}}$ is given by
\bq
\lefteqn{
 \tilde{I}_{2,\mathrm{equal}}\left(q\right)
 =
 3 \ln\left(-q\right)
 } \\
 & &
 + 12 
   \sum\limits_{j=1}^\infty \sum\limits_{k=1}^\infty 
   \frac{1}{j}
    \left[ 6 r_2^j + r_3^j + r_3^{2 j} - 6 r_4^j - 6 r_4^{3 j} + 3 r_6^j + 3 r_6^{5 j} 
           - 2 r_{12}^j - 2 r_{12}^{5 j} - 2 r_{12}^{7 j} - 2 r_{12}^{11 j} \right] q^{j k}.
 \nonumber
\eq
Of particular relevance for the weight $4$ part is the term $3 \ln(-q)$.
We have
\bq
\label{special_solution_equal_mass_case}
\lefteqn{
 -
 3 \frac{\psi_1}{\pi}
 \frac{1}{2i}
 \sum\limits_{j=1}^\infty
 \sum\limits_{k=1}^\infty
 \int\limits_0^q \frac{dq_1}{q_1}
 \int\limits_0^{q_1} \frac{dq_2}{q_2}
 \left(r_3^j - r_3^{-j} \right) 
 k^2 \left(-1\right)^k
 \left(-q_2\right)^{jk}
 \ln\left(-q_2\right)
 = } & &
 \nonumber \\
 & &
 -
 \frac{3}{2i} 
 \frac{\psi_1}{\pi}
 \ln\left(-q\right)
 \left(
 \mathrm{ELi}_{2;0}\left(r_3;-1;-q\right)
 -
 \mathrm{ELi}_{2;0}\left(r_3^{-1};-1;-q\right)
 \right)
 \nonumber \\
 & &
 +
 \frac{3}{i} \frac{\psi_1}{\pi}
 \left(
 \mathrm{ELi}_{3;1}\left(r_3;-1;-q\right)
 -
 \mathrm{ELi}_{3;1}\left(r_3^{-1};-1;-q\right)
 \right).
\eq
Counting the weight of $\ln(-q)$ as two, 
we notice that in eq.~(\ref{special_solution_equal_mass_case}) all terms are of weight $4$. 
The two terms on the right-hand side of eq.~(\ref{special_solution_equal_mass_case}) 
combine nicely with parts of the boundary terms to give
\bq
\label{partial_combination}
 -
 \frac{3}{2} 
 \frac{\psi_1}{\pi}
 \ln\left(-q\right)
 \mathrm{E}_{2;0}\left(r_3;-1;-q\right)
 +
 3 \frac{\psi_1}{\pi}
 \mathrm{E}_{3;1}\left(r_3;-1;-q\right).
\eq
We note that eq.~(\ref{partial_combination}) contains terms of weight $3$ and $4$.
The term proportional to $\ln(-q)$ in eq.~(\ref{partial_combination}) cancels 
the corresponding logarithmic singularity of $\ln(-q)$ (or equivalently $\ln(t)$)
in $F_1(t) S_{111}^{(0)}(2,t)$.
The second term proportional $\mathrm{E}_{3;1}$ explains the occurrence of $\mathrm{E}_{3;1}$ in the final
result for $S_{111}^{(1)}(2,t)$ in eq.~(\ref{res_E_1}).

\section{Conclusions}
\label{sec:conclusions}

In this paper we presented the result for the ${\mathcal O}(\eps^1)$-part
of the sunrise integral around two space-time dimensions.
The result is expressed in terms of generalisations of the Clausen and Glaisher functions towards
the elliptic case.
The ${\mathcal O}(\eps^1)$-part gives us information on elliptic generalisations of multiple polylogarithms
of depth greater than one.
It is worth noting that the ${\mathcal O}(\eps^1)$-part
of the sunrise integral around two space-time dimensions contains terms of weight three and four.
It is not of uniform weight.
We discussed in detail the occurrence of the weight four terms.
Using dimensional-shift relations we expressed the finite part of 
the sunrise integral around four space-time dimensions
in terms
of the ${\mathcal O}(\eps^0)$-part and the ${\mathcal O}(\eps^1)$-part
of the sunrise integral around two space-time dimensions.

\subsection*{Acknowledgements}

C.B. thanks Humboldt University for hospitality and support.

\begin{appendix}

\section{The coefficients of the differential equation}
\label{appendix:coeff}

We first recall from \cite{MullerStach:2011ru} the coefficients of the second-order differential equation
for $S_{111}(2,t)$. This differential equation reads
\bq
 \left[ p_2 \frac{d^2}{d t^2} + p_1 \frac{d}{dt} + p_0 \right] S_{111}\left(2,t\right) & = & \mu^2 p_3,
\eq
where $p_0$, $p_1$ $p_2$ and $p_3$ are given by
\bq
\label{def_p0}
 p_2 & = &
  t \left( t - \mu_1^2 \right)
    \left( t - \mu_2^2 \right)
    \left( t - \mu_3^2 \right)
    \left( t - \mu_4^2 \right)
    \left( 3 t^2 - 2 M_{100} t + \Delta \right),
 \nonumber \\
 p_1 & = & 
  9 t^6
  - 32 M_{100} t^5
  + \left( 37 M_{200} + 70 M_{110} \right) t^4
  - \left( 8 M_{300} + 56 M_{210} + 144 M_{111} \right) t^3
 \\
 & &
  - \left( 13 M_{400} - 36 M_{310} + 46 M_{220} - 124 M_{211} \right) t^2
 \nonumber \\
 & &
  - \left( -8 M_{500} + 24 M_{410} - 16 M_{320} - 96 M_{311} + 144 M_{221} \right) t
 \nonumber \\
 & &
  - \left( M_{600} - 6 M_{510} + 15 M_{420} - 20 M_{330} + 18 M_{411} - 12 M_{321} - 6 M_{222} \right),
 \nonumber \\
 p_0 & = &
  3 t^5
  - 7 M_{100} t^4
  + \left( 2 M_{200} + 16 M_{110} \right) t^3
  + \left( 6 M_{300} - 14 M_{210} \right) t^2
 \nonumber \\
 & &
  - \left( 5 M_{400} - 8 M_{310} + 6 M_{220} - 8 M_{211} \right) t
  + \left( M_{500} - 3 M_{410} + 2 M_{320} + 8 M_{311} - 10 M_{221} \right).
 \nonumber
\eq
The polynomial $p_3$ appearing in the inhomogeneous part is given by
\bq
 p_3 & = & 
 -2 \left( 3 t^2 - 2 M_{100} t + \Delta \right)^2
 \\
 & & 
 + 2 c\left(t,m_1,m_2,m_3\right)  \ln \frac{m_1^2}{\mu^2}
 + 2 c\left(t,m_2,m_3,m_1\right)  \ln \frac{m_2^2}{\mu^2}
 + 2 c\left(t,m_3,m_1,m_2\right)  \ln \frac{m_3^2}{\mu^2},
 \nonumber 
\eq
with
\bq
\lefteqn{
c\left(t,m_1,m_2,m_3\right) = } & &
 \nonumber \\
 & &
 \left( -2 m_1^2 + m_2^2 + m_3^2 \right) t^3
 + \left( 6 m_1^4 - 3 m_2^4 - 3 m_3^4 - 7 m_1^2 m_2^2 - 7 m_1^2 m_3^2 + 14 m_2^2 m_3^2 \right) t^2
 \nonumber \\
 & &
 + \left( -6 m_1^6 + 3 m_2^6 + 3 m_3^6 + 11 m_1^4 m_2^2 + 11 m_1^4 m_3^2 - 8 m_1^2 m_2^4 - 8 m_1^2 m_3^4 - 3 m_2^4 m_3^2 - 3 m_2^2 m_3^4 \right) t
 \nonumber \\
 & & 
 + \left( 2 m_1^8 - m_2^8 - m_3^8 - 5 m_1^6 m_2^2 - 5 m_1^6 m_3^2 + m_1^2 m_2^6 + m_1^2 m_3^6 + 4 m_2^6 m_3^2 + 4 m_2^2 m_3^6 
 \right. \nonumber \\
 & & \left.
        + 3 m_1^4 m_2^4 + 3 m_1^4 m_3^4 - 6 m_2^4 m_3^4 
        + 2 m_1^4 m_2^2 m_3^2 - m_1^2 m_2^4 m_3^2 - m_1^2 m_2^2 m_3^4 \right).
\eq
In $D$ dimensions the integral $S_{111}(D,t)$ satisfies a fourth-order differential equation:
\bq
 \left[
 P_4 \frac{d^4}{dt^4}
 + 
 P_3 \frac{d^3}{dt^3}
 + 
 P_2 \frac{d^2}{dt^2}
 + 
 P_1 \frac{d}{dt}
 + 
 P_0
 \right] S_{111}\left(D,t\right)
 =
 \mu^2
 \left[
 c_{12} T_{12}
 +
 c_{13} T_{13}
 +
 c_{23} T_{23}
 \right].
\eq
The coefficients $P_j$ (with $0 \le j \le 4$) read
\bq
\lefteqn{
 P_4 
 = 
 8 t^3 
 \left( t - \mu_1^2 \right) \left( t - \mu_2^2 \right) \left( t - \mu_3^2 \right) \left( t - \mu_4^2 \right)
 \left[ \left(7-D\right) t^2 - 2 \left(D-3\right) M_{100} t + \left(13-3D\right) \Delta \right],
 } & & 
 \nonumber \\
\lefteqn{
 P_3
 = 
 4 t^2
 \left\{
  \left( 5 D^2 - 71 D + 252 \right) t^6
  - \left( 2 D^2 - 82 D + 500 \right) M_{100} t^5
 \right. 
} & & \nonumber \\
 & & \left.
  - \left( 33 D^2 - 343 D + 520 \right) M_{200} t^4
  - \left( 14 D^2 - 130 D - 144 \right) M_{110} t^4
 \right. \nonumber \\
 & & \left.
  + \left( 52 D^2 - 772 D + 2040 \right) M_{300} t^3
  - \left( 20 D^2 - 100 D + 312 \right) M_{210} t^3
 \right. \nonumber \\
 & & \left.
  - \left( 152 D^2 -2552 D + 9744 \right) M_{111} t^3
  - \left( 13 D^2 - 503 D + 1780 \right) M_{400} t^2
 \right. \nonumber \\
 & & \left.
  + \left( 36 D^2 - 684 D + 2192 \right) M_{310} t^2
  + \left( 124 D^2 - 1332 D + 4592 \right) M_{211} t^2
 \right. \nonumber \\
 & & \left.
  - \left( 46 D^2 - 362 D + 824 \right) M_{220} t^2
  - \left( 18 D^2 + 46 D - 508 \right) M_{500} t
 \right. \nonumber \\
 & & \left.
  + \left( 54 D^2 + 138 D - 1524 \right) M_{410} t
  - \left( 120 D^2 + 904 D - 6288 \right) M_{311} t
 \right. \nonumber \\
 & & \left.
  + \left( 132 D^2 + 1532 D - 9528 \right) M_{221} t
  - \left( 36 D^2 + 92 D - 1016 \right) M_{320} t
 \right. \nonumber \\
 & & \left.
  - 3 D \left( 3 D - 13 \right) \Delta^3
 \right\},
 \nonumber \\
\lefteqn{
 P_2
 = 
 2 t
 \left\{
 - \left( 9 D^3 - 183 D^2 + 1224 D - 2688 \right) t^6
 - \left( 8 D^3 - 62 D^2 - 326 D + 2360 \right) M_{100} t^5
 \right. 
} & &
 \nonumber \\
 & & \left.
 + \left( 51 D^3 - 999 D^2 + 5718 D - 9360 \right) M_{200} t^4
 - \left( 6 D^3 - 78 D^2 + 828 D - 3696 \right) M_{110} t^4
 \right. \nonumber \\
 & & \left.
 - \left( 24 D^3 - 852 D^2 + 6948 D - 15360 \right) M_{300} t^3
 + \left( 56 D^3 - 852 D^2 + 4516 D - 8928 \right) M_{210} t^3
 \right. \nonumber \\
 & & \left.
 + \left( 192 D^3 - 4392 D^2 + 31560 D - 71856 \right) M_{111} t^3
 - \left( 19 D^3 - 73 D^2 - 1636 D + 6640 \right) M_{400} t^2
 \right. \nonumber \\
 & & \left.
 + \left( 12 D^3 + 28 D^2 - 2304 D + 8528 \right) M_{310} t^2
 + \left( 20 D^3 + 772 D^2 - 9536 D + 27248 \right) M_{211} t^2
 \right. \nonumber \\
 & & \left.
 + \left( 14 D^3 - 202 D^2 + 1336 D - 3776 \right) M_{220} t^2
 - \left( 114 D^2 - 414 D - 312 \right) M_{500} t
 \right. \nonumber \\
 & & \left.
 + \left( 342 D^2 - 1242 D - 936 \right) M_{410} t
 + \left( 144 D^3 - 2472 D^2 + 7608 D + 1872 \right) M_{311} t
 \right. \nonumber \\
 & & \left.
 - \left( 288 D^3 - 4260 D^2 + 12732 D + 1872 \right) M_{221} t
 - \left( 228 D^2 - 828 D - 624 \right) M_{320} t
 \right. \nonumber \\
 & & \left.
 - 3 D \left(D-2\right) \left(3D-13\right) \Delta^3
 \right\},
 \nonumber \\
\lefteqn{
 P_1
 = 
 \left( 7 D^4 - 175 D^3 + 1610 D^2 - 6440 D + 9408 \right) t^6
} & &
 \nonumber \\
 & &
 + \left( 14 D^4 - 270 D^3 + 1708 D^2 - 3792 D + 1440 \right) M_{100} t^5
 \nonumber \\
 & &
 - \left( 27 D^4 - 767 D^3 + 7182 D^2 - 27352 D + 36480 \right) M_{200} t^4
 \nonumber \\
 & &
 + \left( 38 D^4 - 782 D^3 + 6172 D^2 - 22384 D + 31296 \right) M_{110} t^4
 \nonumber \\
 & &
 - \left( 12 D^4 + 52 D^3 - 2712 D^2 + 16688 D - 29760 \right) M_{300} t^3
 \nonumber \\
 & &
 - \left( 20 D^4 - 532 D^3 + 4888 D^2 - 19376 D + 28608 \right) M_{210} t^3
 \nonumber \\
 & &
 - \left( 56 D^4 - 2168 D^3 + 24688 D^2 - 112576 D + 179136 \right) M_{111} t^3
 \nonumber \\
 & &
 + \left( 17 D^4 - 273 D^3 + 1342 D^2 - 1416 D - 2880 \right) M_{400} t^2
 \nonumber \\
 & &
 - \left( 20 D^4 - 180 D^3 + 184 D^2 + 2592 D - 7488 \right) M_{310} t^2
 \nonumber \\
 & &
 - \left( 140 D^4 - 2220 D^3 + 11656 D^2 - 19680 D - 1728 \right) M_{211} t^2
 \nonumber \\
 & &
 + \left( 6 D^4 + 186 D^3 - 2316 D^2 + 8016 D - 9216 \right) M_{220} t^2
 \nonumber \\
 & &
 - \left( 2 D^4 - 34 D^3 + 292 D^2 - 1088 D + 1248 \right) M_{500} t
 \nonumber \\
 & &
 + \left( 6 D^4 - 102 D^3 + 876 D^2 - 3264 D + 3744 \right) M_{410} t
 \nonumber \\
 & &
 + \left( 104 D^4 - 1128 D^3 + 2992 D^2 + 1440 D - 7488 \right) M_{311} t
 \nonumber \\
 & &
 - \left( 220 D^4 - 2460 D^3 + 7736 D^2 - 3648 D - 7488 \right) M_{221} t
 \nonumber \\
 & &
 - \left( 4 D^4 - 68 D^3 + 584 D^2 - 2176 D + 2496 \right) M_{320} t
 \nonumber \\
 & &
 - D \left(D-2\right) \left(D-4\right) \left(3D-13\right) \Delta^3,
 \nonumber \\
\lefteqn{
 P_0 = 
 \left(D-3\right) \left(D-4\right)
 \left\{
 - \left( D^3 - 21 D^2 + 146 D - 336 \right) t^5
 - \left( 3 D^3 - 51 D^2 + 258 D - 360 \right) M_{100} t^4
 \right. 
} & &
\nonumber \\
 & & \left.
 + \left( 2 D^3 - 70 D^2 + 568 D - 1280 \right) M_{200} t^3
 - \left( 12 D^3 - 204 D^2 + 1144 D - 2192 \right) M_{110} t^3
 \right. \nonumber \\
 & & \left.
 + \left( 6 D^3 - 66 D^2 + 120 D + 240 \right) M_{300} t^2
 - \left( 6 D^3 - 50 D^2 + 40 D + 336 \right) M_{210} t^2
 \right. \nonumber \\
 & & \left.
 - \left( 28 D^3 - 300 D^2 + 584 D + 1392 \right) M_{111} t^2 
 - \left( D^3 - 33 D^2 + 182 D - 240 \right) M_{400} t
 \right. \nonumber \\
 & & \left.
 + \left( 12 D^3 - 148 D^2 + 560 D - 624 \right) M_{310} t
 - \left( 12 D^3 - 84 D^2 + 48 D + 144 \right) M_{211} t
 \right. \nonumber \\
 & & \left.
 - \left( 22 D^3 - 230 D^2 + 756 D - 768 \right) M_{220} t
 \right. \nonumber \\
 & & \left.
 + \left(D-2\right) \left(3D-13\right)
   \left[ \left(D-4\right) M_{300} - \left(D-4\right) M_{210} + \left(2D-16\right) M_{111} \right]
   \Delta
 \right\}.
\eq
In the inhomogeneous term we have
\bq
\lefteqn{
 c_{12} = 
 \left( D-2 \right)^2 \left(D-4\right)
 \left\{
  \left( D^2 - 13 D + 42 \right) t^4
 \right. } & & \nonumber \\
 & & \left.
 + D^2 \left( 2 m_1^2 + 2 m_2^2 + 4 m_3^2 \right) t^3
 - D \left( 24 m_1^2 + 24 m_2^2 + 40 m_3^2 \right) t^3
 + \left( 62 m_1^2 + 62 m_2^2 + 76 m_3^2 \right) t^3
 \right. \nonumber \\
 & & \left.
 + D^2 \left(-4 m_1^4 -4 m_2^4 + 2 m_3^4 + 10 m_1^2 m_3^2 + 10 m_2^2 m_3^2 + 8 m_1^2 m_2^2 \right) t^2
 \right. \nonumber \\
 & & \left.
 + D \left( 62 m_1^4 + 62 m_2^4 + 22 m_3^4 - 132 m_1^2 m_3^2 - 132 m_2^2 m_3^2 - 108 m_1^2 m_2^2 \right) t^2
 \right. \nonumber \\
 & & \left.
 + \left(-198 m_1^4 -198 m_2^4 -180 m_3^4 + 450 m_1^2 m_3^2 + 450 m_2^2 m_3^2 + 348 m_1^2 m_2^2 \right) t^2
 \right. \nonumber \\
 & & \left.
 + D^2 \left( -2 m_1^6 -2 m_2^6 -4 m_3^6 + 2 m_1^4 m_2^2 + 2 m_1^2 m_2^4 -8 m_1^4 m_3^2 -8  m_2^4 m_3^2 + 14 m_1^2 m_3^4 + 14 m_2^2 m_3^4 
 \right. \right. \nonumber \\
 & & \left. \left.
              + 16 m_1^2 m_2^2 m_3^2 \right) t
 \right. \nonumber \\
 & & \left.
 + 32 D m_3^2  \left( 3 m_1^4 +3 m_2^4 + 2 m_3^4 -5 m_1^2 m_3^2 -5 m_2^2 m_3^2 -4 m_1^2 m_2^2 \right) t
 \right. \nonumber \\
 & & \left.
 + \left( 42 m_1^6 + 42 m_2^6 -180 m_3^6 -42 m_1^4 m_2^2 -42 m_1^2 m_2^4 -288 m_1^4 m_3^2 -288 m_2^4 m_3^2 + 426 m_1^2 m_3^4 
 \right. \right. \nonumber \\
 & & \left. \left.
          + 426 m_2^2 m_3^4 
          + 192 m_1^2 m_2^2 m_3^2 \right) t
 \right. \nonumber \\
 & & \left.
 + \left( 3 D -13 \right)
   \left[
          D \left( -m_1^4-m_2^4 + m_3^4 + 2 m_1^2 m_2^2 \right) + 4 m_1^4 + 4 m_2^4 - 6 m_3^4 + 2 m_1^2 m_3^2 + 2 m_2^2 m_3^2 
 \right. \right. \nonumber \\
 & & \left. \left.
    - 8 m_1^2 m_2^2 
   \right]
   \Delta
 \right\}.
\eq
$c_{13}$ and $c_{23}$ are obtained by permutation of the masses.

\section{The integration constants}
\label{section_integration_constants}

The explicit expressions for the two integration constants $C_1$ and $C_2$ appearing in eq.~(\ref{L2_S_111_1}) are given by
\bq
 C_1
 & = &
 - \frac{4}{3} \left(  M_{300} - M_{210} + 5 M_{111} \right) S_{111}^{(0)}\left(2,0\right)
 + \mu^2 \left[
 \frac{2}{3} \Delta
 \right. \nonumber \\
 & &
 + \frac{2}{3} \left( 4 m_1^4 - m_2^4 - m_3^4 - 3 m_1^2 m_2^2 - 3 m_1^2 m_3^2 + 2 m_2^2 m_3^2 \right) \ln\left(\frac{m_1^2}{\mu^2}\right)
 \nonumber \\
 & &
 + \frac{2}{3} \left( 4 m_2^4 - m_3^4 - m_1^4 - 3 m_2^2 m_3^2 - 3 m_2^2 m_1^2 + 2 m_3^2 m_1^2 \right) \ln\left(\frac{m_2^2}{\mu^2}\right)
 \nonumber \\
 & &
 + \frac{2}{3} \left( 4 m_3^4 - m_1^4 - m_2^4 - 3 m_3^2 m_1^2 - 3 m_3^2 m_2^2 + 2 m_1^2 m_2^2 \right) \ln\left(\frac{m_3^2}{\mu^2}\right)
 \nonumber \\
 & &
 - \frac{1}{3} \left( 2 m_1^4 - m_2^4 - m_3^4 - m_1^2 m_2^2 - m_1^2 m_3^2 + 2 m_2^2 m_3^2 \right) \ln^2\left(\frac{m_1^2}{\mu^2}\right)
 \nonumber \\
 & &
 - \frac{1}{3} \left( 2 m_2^4 - m_3^4 - m_1^4 - m_2^2 m_3^2 - m_2^2 m_1^2 + 2 m_3^2 m_1^2 \right) \ln^2\left(\frac{m_2^2}{\mu^2}\right)
 \nonumber \\
 & &
 - \frac{1}{3} \left( 2 m_3^4 - m_1^4 - m_2^4 - m_3^2 m_1^2 - m_3^2 m_2^2 + 2 m_1^2 m_2^2 \right) \ln^2\left(\frac{m_3^2}{\mu^2}\right)
 \nonumber \\
 & &
 - \frac{2}{3} \left( m_1^4 + m_2^4 - 2 m_3^4 + m_1^2 m_3^2 + m_2^2 m_3^2 - 2 m_1^2 m_2^2 \right) \ln\left(\frac{m_1^2}{\mu^2}\right) \ln\left(\frac{m_2^2}{\mu^2}\right)
 \nonumber \\
 & &
 - \frac{2}{3} \left( m_2^4 + m_3^4 - 2 m_1^4 + m_2^2 m_1^2 + m_3^2 m_1^2 - 2 m_2^2 m_3^2 \right) \ln\left(\frac{m_2^2}{\mu^2}\right) \ln\left(\frac{m_3^2}{\mu^2}\right)
 \nonumber \\
 & & \left.
 - \frac{2}{3} \left( m_3^4 + m_1^4 - 2 m_2^4 + m_3^2 m_2^2 + m_1^2 m_2^2 - 2 m_3^2 m_1^2 \right) \ln\left(\frac{m_3^2}{\mu^2}\right) \ln\left(\frac{m_1^2}{\mu^2}\right)
 \right],
\eq
\bq
\lefteqn{
 C_2 = 
 \frac{4}{7 \Delta^2} \left( 7 M_{600} - 44 M_{510} + 113 M_{420} - 77 M_{411} + 121 M_{321} - 152 M_{330} - 258 M_{222} \right) S_{111}^{(0)}\left(2,0\right)
 } & &
 \nonumber \\
 & &
 + \mu^2 \left\{
 \frac{1}{7 \Delta} \left( 121 M_{300} - 121 M_{210} + 270 M_{111} \right)
 \right. \nonumber \\
 & &
 + \left(2 m_1^2 - m_2^2 - m_3^2 \right) \ln^2\left(\frac{m_1^2}{\mu^2} \right)
 + \left(2 m_2^2 - m_3^2 - m_1^2 \right) \ln^2\left(\frac{m_2^2}{\mu^2} \right)
 + \left(2 m_3^2 - m_1^2 - m_2^2 \right) \ln^2\left(\frac{m_3^2}{\mu^2} \right)
 \nonumber \\
 & &
 + 2 \left( m_1^2 + m_2^2 - 2 m_3^2 \right) \ln\left(\frac{m_1^2}{\mu^2} \right) \ln\left(\frac{m_2^2}{\mu^2} \right)
 + 2 \left( m_2^2 + m_3^2 - 2 m_1^2 \right) \ln\left(\frac{m_2^2}{\mu^2} \right) \ln\left(\frac{m_3^2}{\mu^2} \right)
 \nonumber \\
 & &
 + 2 \left( m_3^2 + m_1^2 - 2 m_2^2 \right) \ln\left(\frac{m_3^2}{\mu^2} \right) \ln\left(\frac{m_1^2}{\mu^2} \right)
 \nonumber \\
 & &
 + \frac{2}{7 \Delta^2} 
   \left[ 
          32 m_1^{10} 
          + 55 \left(m_2^2+m_3^2 \right) m_1^8
          - \left( 355 m_2^4 + 355 m_3^4 - 400 m_2^2 m_3^2 \right) m_1^6
 \right. \nonumber \\
 & & \left.
          + 5 \left(m_2^2+m_3^2\right) \left( 83 m_2^4 + 83 m_3^4 - 124 m_2^2 m_3^2 \right) m_1^4
          - \left( m_2^2 - m_3^2 \right)^2 \left( 145 m_2^4 + 145 m_3^4 + 546 m_2^2 m_3^2 \right) m_1^2
 \right. \nonumber \\
 & & \left.
          -2 \left( m_2^2 - m_3^2 \right)^4 \left( m_2^2 + m_3^2 \right)
   \right] \ln\left(\frac{m_1^2}{\mu^2} \right)
 \nonumber \\
 & &
 + \frac{2}{7 \Delta^2} 
   \left[ 
          32 m_2^{10} 
          + 55 \left(m_3^2+m_1^2 \right) m_2^8
          - \left( 355 m_3^4 + 355 m_1^4 - 400 m_3^2 m_1^2 \right) m_2^6
 \right. \nonumber \\
 & & \left.
          + 5 \left(m_3^2+m_1^2\right) \left( 83 m_3^4 + 83 m_1^4 - 124 m_3^2 m_1^2 \right) m_2^4
          - \left( m_3^2 - m_1^2 \right)^2 \left( 145 m_3^4 + 145 m_1^4 + 546 m_3^2 m_1^2 \right) m_2^2
 \right. \nonumber \\
 & & \left.
          -2 \left( m_3^2 - m_1^2 \right)^4 \left( m_3^2 + m_1^2 \right)
   \right] \ln\left(\frac{m_2^2}{\mu^2} \right)
 \nonumber \\
 & &
 + \frac{2}{7 \Delta^2} 
   \left[ 
          32 m_3^{10} 
          + 55 \left(m_1^2+m_2^2 \right) m_3^8
          - \left( 355 m_1^4 + 355 m_2^4 - 400 m_1^2 m_2^2 \right) m_3^6
 \right. \nonumber \\
 & & \left.
          + 5 \left(m_1^2+m_2^2\right) \left( 83 m_1^4 + 83 m_2^4 - 124 m_1^2 m_2^2 \right) m_3^4
          - \left( m_1^2 - m_2^2 \right)^2 \left( 145 m_1^4 + 145 m_2^4 + 546 m_1^2 m_2^2 \right) m_3^2
 \right. \nonumber \\
 & & \left. \left.
          -2 \left( m_1^2 - m_2^2 \right)^4 \left( m_1^2 + m_2^2 \right)
   \right] \ln\left(\frac{m_3^2}{\mu^2} \right)
 \right\}.
\eq

\end{appendix}

\bibliography{/home/stefanw/notes/biblio}
\bibliographystyle{/home/stefanw/latex-style/h-physrev5}

\end{document}